\documentclass[11pt,a4paper]{article}

\usepackage{jheppub}

\usepackage{amssymb}
\usepackage{amsfonts}
\usepackage{graphicx}
\usepackage{epstopdf}
\usepackage{dcolumn}
\usepackage{amsmath}
\usepackage{latexsym,bm}
\usepackage{amsthm}
\usepackage{slashed}
\usepackage{float}
\usepackage{color}
\usepackage{url}

\def \be {\begin{equation}}
\def \ee {\end{equation}}
\def \bsp {\begin{split}}
\def \esp {\end{split}}
\def \bea {\begin{eqnarray}}
\def \eea {\end{eqnarray}}

\def\mc{\mathcal}

\def\O{\mathcal{O}}

\newcommand{\fracs}[2]{{\textstyle{#1\over #2}}}

\title{Tuned and Non-Higgsable U(1)s in F-theory}

\author[a]{Yi-Nan Wang}

\affiliation[a]{Center for Theoretical Physics,\\Department of Physics\\Massachusetts Institute of Technology\\77 Massachusetts Avenue\\Cambridge, MA 02139, USA}

\emailAdd{wangyn@mit.edu}

\preprint{\today \hspace*{0.1in} MIT-CTP-4856}

\abstract{We study the tuning of U(1) gauge fields in F-theory models on a base of general dimension. We construct a formula that computes the change in Weierstrass moduli when such a U(1) is tuned, based on the Morrison-Park form of a Weierstrass model with an additional rational section. Using this formula, we propose the form of ``minimal tuning'' on any base, which corresponds to the case where the decrease in the number of Weierstrass moduli is minimal. Applying this result, we discover some universal features of bases with non-Higgsable U(1)s. Mathematically, a generic elliptic fibration over such a base has additional rational sections. Physically, this condition implies the existence of U(1) gauge group in the low-energy supergravity theory after compactification that cannot be Higgsed away. In particular, we show that the elliptic Calabi-Yau manifold over such a base has a small number of complex structure moduli. We also suggest that non-Higgsable U(1)s can never appear on any toric bases. Finally, we construct the first example of a threefold base with non-Higgsable U(1)s.
 }


\begin{document}

\maketitle

\flushbottom

\section{Introduction}

F-theory \cite{F-theory, Morrison-Vafa-I, Morrison-Vafa-II} is a powerful geometric description of non-perturbative type IIB superstring theory. The IIB string theory is compactified on a complex $d$-fold $B$ to construct supergravity theories in $(10-2d)$ real dimensions. The monodromy information of the axiodilaton is encoded in an elliptically fibered Calabi-Yau $(d+1)$-fold $X$ over the base $B$. There can be a base locus on which the elliptic fiber is degenerate. In the IIB language, this corresponds to seven-brane configurations carrying gauge groups and matter in the low-energy supergravity theory. The advantage of F-theory is two-fold. First, the base $B$ itself is not necessarily Calabi-Yau, hence F-theory provides a much richer geometric playground. Second, F-theory effectively incorporates exceptional gauge groups that are hard to construct in weakly coupled IIB string theory, such as $G_2$, $E_6$ and $E_8$. 

However, the simplest gauge group, U(1), is tricky to describe with F-theory geometric techniques. While the non-Abelian gauge groups can be read off from local information on the base, the Abelian gauge groups are related to additional rational sections on the elliptic manifold $X$. Mathematically, these rational sections form the Mordell-Weil group of the elliptic fibration, which is generally hard to construct and analyze.

In a recent work by Morrison and Park \cite{Morrison-Park}, a generic Weierstrass form with non-zero Mordell-Weil rank was proposed (\ref{mp}). It is parameterized by sections of various line bundles. The only variable characterizing the way of tuning U(1) on a given base is a line bundle $L$ over the base. Further investigation and generalization of this form are carried out in \cite{MTsection,Klevers-WT,TallSection}. Additionally, an elliptic fibration with multiple sections is a useful geometric setup in F-theory GUT model building \cite{Mayrhofer:2012zy,Braun:2013nqa,Cvetic:2013uta,Borchmann:2013hta,Cvetic:2013jta,Cvetic:2013nia,Cvetic:2013qsa,Lawrie:2015hia}.

However, the counting of independent degrees of freedom in the Morrison-Park form is not obvious, since the parameters in the  Morrison-Park form may include redundant components. The first goal of this paper is to construct a formula that counts the decrease in Weierstrass moduli after a U(1) is tuned. We start from the generic fibration over the base $B$, which corresponds to the ``non-Higgsable'' phase \cite{clusters,4D-NHC} and the gauge groups in the supergravity are minimal. We propose such a formula (\ref{dh21g}) in Section~\ref{s:counting}. We argue that this formula is exact if the line bundle $L$ is base point free. In other cases, the formula (\ref{dh21g}) serves as a universal lower bound.

With this formula, we conjecture that the ``minimal'' tuning of U(1) on a base is given by the Morrison-Park form with the choice $L=0$. We have checked that this result holds for many toric 2D bases. Moreover, we proved that this choice is indeed minimal for generalized del Pezzo surfaces \cite{Derenthal,DerenthalThesis} using anomaly arguments.

Furthermore, if we can compute the decrease in Weierstrass moduli in the minimal tuning, then we can exactly identify the criterion for a non-Higgsable U(1) on a base $B$. That is, if the required number of tuned Weierstrass moduli is non-positive, then we know that there is a non-Higgsable U(1) before any tuning from the non-Higgsable phase. We prove a set of constraints on the Weierstrass polynomials $f$ and $g$ for a base with this property in Section~\ref{s:constraints}, assuming the form of minimal tuning. We find out that Newton polytopes of these polynomials have to align along a 1D line, and the maximal number of monomials in $f$ and $g$ are constrained to be 5 and 7. Moreover, we show that there can never be a non-Higgsable U(1) on any toric base in arbitrary dimension, which is consistent with the results in \cite{mt-toric,Hodge}. We also conjecture that the general form of a $d$-dimensional base with non-Higgsable U(1)s is a resolution of a Calabi-Yau $(d-1)$-fold fibration over $\mathbb{P}^1$. For the case of $d=2$, this is reduced to the generalized Schoen constructions in \cite{FiberProduct}. For the case of $d=3$, this suggests that the bases with non-Higgsable U(1)s are a resolution of K3 or $T^4$ fibrations over $\mathbb{P}^1$.

As a check, we compute the Weierstrass polynomials for all the semi-toric 2D bases with non-Higgsable U(1)s, listed in  \cite{Martini-WT}. We find that the Newton polytopes of Weierstrass polynomials indeed fit in our bound.

More interestingly, we construct the first example of a 3D base with non-Higgsable U(1)s in Section~\ref{s:3D}. This threefold geometry indeed contains degenerate K3 fibers. Finally, some conclusions and further directions are shown in Section~\ref{s:conclusion}.

\section{Minimal tuning of U(1) on a general base}

\subsection{The Morrison-Park form for an additonal rational section}\label{s:mp}

We start from a generic complex $d$-dimensional base $B$ with anticanonical line bundle (class) $-K$. A generic elliptically fibered Calabi-Yau manifold $X$ over $B$ can be written in the following Weierstrass form:
\be
y^2=x^3+fx+g.
\ee
$f\in\mathcal{O}(-4K)$, $g\in\mathcal{O}(-6K)$ are generic sections of line bundles $-4K$ and $-6K$ over the base $B$. $f$ and $g$ are ``generic'' in the sense that the coefficients of the monomials (base vectors for the linear system $|-4K|$, $|-6K|$) are randomly chosen, so that $X$ has the most complex structure moduli (largest $h^{d,1}(X)$). In the F-theory context, this case corresponds to the ``non-Higgsable'' phase \cite{clusters,4D-NHC}, where the gauge groups in the lower dimensional supergravity are minimal. These minimal gauge groups are called non-Higgsable gauge groups. For non-Abelian gauge groups, they correspond to local structures on the base $B$, which are called non-Higgsable clusters (NHC). They appear when $f$, $g$ and the discriminant $\Delta=4f^3+27g^2$ vanish to certain degrees on some divisors.

For a good base $B$ in F-theory, we require that the anticanonical line bundle $-K$ is effective, otherwise these sections $f$ and $g$ do not exist. Additionally, the degree of vanishing for $f$ and $g$ cannot reach $(4,6)$ at a codimension-one or codimension-two locus on the base $B$. For the latter case, the codimension-two locus on the base $B$ has to be blown up until we get a good base without these locus. In the context of F-theory construction of 6D (1,0) superconformal field theory, this corresponds to moving into the tensor branch \cite{Heckman:2013pva,DelZotto:2014hpa,Heckman:2015bfa}.

$h^{d,1}(X)$ of the elliptic Calabi-Yau manifold can be written as
\be
h^{d,1}(X)=W-\omega_{\mathrm{aut}}+N_{nW}-1.\label{hd1}
\ee
Here 
\be
W=h^0(-4K)+h^0(-6K)
\ee
is the number of Weierstrass moduli, or the total number of monomials in $f$ and $g$. $\omega_{\mathrm{aut}}$ is the dimension of the automorphism group of the base. $N_{nW}$ is some other non-Weierstrass contributions. For example, in the $d=2$ case, $N_{nW}=N_{-2}$, the number of (-2)-curves on the base that are not in any non-Higgsable clusters \cite{mt-toric}.

$\omega_{\mathrm{aut}}$ is determined by the properties of the base $B$, so it does not change when we tune a gauge group on $B$. $N_{nW}$ does change in some rare cases. Taking a $d=2$ example, if the degree of vanishing of the discriminant $\Delta$ goes from 0 to some positive number on a (-2)-curve, then this (-2)-curve is no longer counted in the term $N_{-2}$ \cite{Johnson-WT}. Hence the decrease in $h^{d,1}(X)$ is generally
\be
-\Delta h^{d,1}(X)=-\Delta W-\Delta N_{nW}.
\ee
In this paper, we never encounter any example where $\Delta_{nW}\neq 0$.

Now we want to have a U(1) on $B$. Abelian groups in F-theory are special in the sense that they correspond to additional global rational sections on $B$. The generic Weierstrass form after the U(1) is tuned can be written in the following Morrison-Park form \cite{Morrison-Park}\footnote{Weierstrass models with a U(1) but not in the Morrison-Park form has been found \cite{Klevers-WT}. It is shown that this exotic construction can be reduced to a non-Calabi-Yau Morrison-Park form after a birational transformation \cite{TallSection}. The generic form of such constructions is not known, and we will not consider these cases in this paper. The Morrison-Park form can describe F-theory models with multiple U(1)s as well, for example the tuning of U(1)$\times$U(1) in Section~\ref{s:P2} and multiple non-Higgsable U(1)s in Section~\ref{s:2D}.}:
\be
y^2=x^3+(c_1 c_3-\frac{1}{3}c_2^2-b^2 c_0)xz^4+(c_0 c_3^2+\frac{2}{27}c_2^3-\frac{1}{3}c_1 c_2 c_3-\frac{2}{3}b^2 c_0 c_2+\frac{1}{4}b^2 c_1^2)z^6. \label{mp}
\ee

The coefficients $b$, $c_0$, $c_1$, $c_2$ and $c_3$ are holomorphic sections of line bundles $L$, $-4K-2L$, $-3K-L$, $-2K$ and $-K+L$ on $B$ respectively:
\be
\bsp
&b\in\mathcal{O}(L)\\
&c_0\in\mathcal{O}(-4K-2L)\\
&c_1\in\mathcal{O}(-3K-L)\\
&c_2\in\mathcal{O}(-2K)\\
&c_3\in\mathcal{O}(-K+L)\label{sections}
\end{split}
\ee

The rational section over $B$ is then given by:
\be
(x,y,z)=(\lambda,\alpha,b)=(c_3^2-\frac{2}{3}b^2 c_2,-c_3^3+b^2 c_2 c_3-\frac{1}{2}b^4 c_1,b)\label{rationals}
\ee
Apparently
\be
\bsp
&\lambda\in\mathcal{O}(-2K+2L)\\
&\alpha\in\mathcal{O}(-3K+3L).
\end{split}
\ee

Hence $L$ is an effective bundle on $B$ with holomorphic section $b$, which characterizes the particular way of tuning the U(1). For some bases with non-Higgsable clusters of high rank gauge groups, taking $L=0$ may lead to non-minimal singularities in the total space $X$ that cannot be resolved \cite{MTsection}. We will discuss this issue explicitly for $B=\mathbb{F}_{12}$ in Section (\ref{s:F12}).

For $c_0$ and $c_1$, it is not clear whether their holomorphic sections exist. If the line bundle $-2K-L$ is effective, which is denoted by $-2K-L\geq 0$ or equivalently $L\leq -2K$, then $c_0$ and $c_1$ both have holomorphic sections. In \cite{Morrison-Park}, there is an extension of (\ref{sections}) when $-2K-L$ is not effective. In this case, we just take $c_0\equiv 0$. 

Now the discriminant of the Weierstrass form (\ref{mp}) is
\be
\Delta=\frac{27}{16}b^4 c_1^4+b^2 c_1^2 c_2^3-\frac{9}{2}b^2 c_1^3 c_2 c_3-c_1^2 c_2^2 c_3^2+4 c_1^3 c_3^3,
\ee
which means that an SU(2) gauge group exists on the curve $c_1=0$.

In this paper, we use the weaker constraint $-3K-L\geq 0$, or $L\leq -3K$. 

Note that this condition $L\leq -3K$ cannot be further relaxed, otherwise $c_0=c_1=0$, and the Weierstrass form (\ref{mp}) becomes
\be
y^2=x^3-\frac{1}{3}c_2^2 xz^4+\frac{2}{27}c_2^3 z^6,
\ee
which is globally singular over the base $B$.

\subsection{Counting independent variables}\label{s:counting}

A previously unsolved issue when using the Morrison-Park form (\ref{mp}) is that the functions $b$, $c_0$, $c_1$, $c_2$ and $c_3$ are not independent. There may exist an infinitesimal transformation: $b\rightarrow b+\delta b$, $c_0\rightarrow c_0+\delta c_0$, $c_1\rightarrow c_1+\delta c_1$, $c_2\rightarrow c_2+\delta c_2$, $c_3\rightarrow c_3+\delta c_3$ such that the rational points $\frac{\lambda}{b^2}$, $\frac{\alpha}{b^3}$ and $f$, $g$ are invariant\footnote{Similar redundancy issue also appears in the tuning of SU(7) gauge group\cite{Anderson:2015cqy}.}. Actually, because of the relation
\be
\frac{\alpha^2}{b^6}=\frac{\lambda^3}{b^6}+\frac{f\lambda}{b^2}+g,
\ee
if $\frac{\lambda}{b^2}$, $\frac{\alpha}{b^3}$ and $f$ are fixed, then $g$ is also fixed. So we only need to guarantee the invariance of $\frac{\lambda}{b^2}$, $\frac{\alpha}{b^3}$ and $f$.

The number of such infinitesimal transformations then gives the number of redundant variables in the Morrison-Park form, $N_r$.

If we can compute this number $N_r$, then the number of Weierstrass moduli in the Morrison-Park form equals to
\be
W'=h^0(L)+h^0(-4K-2L)+h^0(-3K-L)+h^0(-2K)+h^0(-K+L)-N_r.
\ee
Before the U(1) is tuned, the number of Weierstrass moduli in the non-Higgsable phase equals to
\be
W=h^0(-4K)+h^0(-6K).
\ee
Hence the number of tuned Weierstrass moduli equals to
\be
\bsp
W-W'=&h^0(-4K)+h^0(-6K)-h^0(L)-h^0(-4K-2L)-h^0(-3K-L)-h^0(-2K)\\
&-h^0(-K+L)+N_r.
\end{split}
\ee

We know that $N_r\geq 1$, since there is a trivial rescaling automorphism: $b\rightarrow tb$, $c_3\rightarrow tc_3$, $c_2\rightarrow c_2$, $c_1\rightarrow t^{-1}c_1$, $c_0\rightarrow t^{-2}c_0$ that keeps $\frac{\lambda}{b^2}$, $\frac{\alpha}{b^3}$, $f$ and $g$ invariant. But other transformations may exist as well.
 
Now we study the simplest case $L=0$ first. Since we have already taken the rescaling automorphism into account, we can set $b=1$ for simplicity. Now the Morrison-Park form becomes
\be
y^2=x^3+(c_1 c_3-\frac{1}{3}c_2^2- c_0)xz^4+(c_0 c_3^2+\frac{2}{27}c_2^3-\frac{1}{3}c_1 c_2 c_3-\frac{2}{3} c_0 c_2+\frac{1}{4}c_1^2)z^6.
\ee
The rational section is
\be
(x,y,z)=(\lambda,\alpha,b)=(c_3^2-\frac{2}{3}c_2,-c_3^3+c_2 c_3-\frac{1}{2}c_1,1).
\ee
If $\delta\lambda=0$ under an infinitesimal transformation, then it is required that
\be
\delta c_2=3c_3 \delta c_3.
\ee
Plugging this equation into the requirement $\delta\alpha=0$, we derive
\be
\delta c_1=2c_2 \delta c_3.
\ee
Then the explicit form of $\delta f=0$ tells us
\be
\delta c_0=c_1 \delta c_3.
\ee

Now one can easily check that under
\be
\bsp
\delta c_2&=3c_3 \delta c_3,\\
\delta c_1&=2c_2 \delta c_3,\\
\delta c_0&=c_1 \delta c_3,
\end{split}
\ee
$g$ is invariant.

This infinitesimal transformation is parametrized by an arbitrary section $\delta c_3\in\O(-K)$, which implies that the coefficient $c_3$ is actually a dummy variable. Hence the total number of redundant variables equals to
\be
N_r=1+h^0(-K).
\ee
We have thus derived the formula for $(-\Delta W)$ in the case of $L=0$:
\be
-\Delta W_{L=0}=h^0(-6K)-h^0(-3K)-h^0(-2K)\label{dh21L0}.
\ee

Now we study the more general case $L>0$. Similarly, the infinitesimal transformations parametrized by $\delta c_3$ and $\delta b$ which leave $\frac{\lambda}{b^2}$, $\frac{\alpha}{b^3}$, $f$ and $g$ invariant are in the following form:
\be
\bsp
\delta c_2&=\frac{3c_3 \delta c_3}{b^2}-\frac{3c_3^2\delta b}{b^3},\\
\delta c_1&=\frac{2c_2 \delta c_3}{b^2}-\frac{c_1\delta b}{b}-\frac{2c_2 c_3\delta b}{b^3},\\
\delta c_0&=\frac{c_1 \delta c_3}{b^2}-\frac{c_1 c_3\delta b}{b^3}-\frac{2 c_0\delta b}{b}.\label{gtrans}
\end{split}
\ee
However, they are rational functions rather than holomorphic functions. Hence these infinitesimal transformations are not apparently legitimate. Nevertheless, we are guaranteed to have a subset of infinitesimal transformations:
\be
\delta b\equiv 0\ ,\ b^2|\delta c_3,\label{freetrans}
\ee
which indeed give holomorphic $\delta c_2$, $\delta c_1$ and $\delta c_0$. The condition $b^2|\delta c_3$ tells us that $\delta c_3'=\delta c_3/b^2$ is a holomorphic section of the line bundle $\O(-K-L)$. Hence we obtain a lower bound on $N_r$:
\be
N_r\geq 1+h^0(-K-L).
\ee

We have thus derived a lower bound for $(-\Delta W)$ for general $L$:
\be
\bsp
-\Delta W\geq &h^0(-4K)+h^0(-6K)-h^0(L)-h^0(-4K-2L)-h^0(-3K-L)-h^0(-2K)\\
&-h^0(-K+L)+h^0(-K-L)+1\label{dh21gineq}.
\end{split}
\ee

For base point free line bundles $L$ (there is no base point $x_0$ on which every section $s\in\mc{O}(L)$ satisfies $s(x_0)=0$), we claim that the above inequality is saturated, so we get the exact formula:
\be
\bsp
-\Delta W=&h^0(-4K)+h^0(-6K)-h^0(L)-h^0(-4K-2L)-h^0(-3K-L)-h^0(-2K)\\
&-h^0(-K+L)+h^0(-K-L)+1\label{dh21g}.
\end{split}
\ee

Because $L$ has no base point, for generic sections $c_3\in\mc{O}(-K+L)$ and $b\in\mc{O}(L)$, they do not share any common factors after these polynomials are factorized to irreducible components (if they have). For $\delta c_2$ in (\ref{gtrans}) to be holomorphic:
\be
\delta c_2=\frac{3c_3 b\delta c_3-3c_3^2\delta b}{b^3},
\ee
it is required that $\delta b=tb$ where $t$ is a complex number. But this form of $\delta b$ is exactly the trivial rescaling isomorphism, so we want to substract this component and set $\delta b=0$. Now
\be
\delta c_2=\frac{3c_3\delta c_3}{b^2}.
\ee
Because $c_3$ does not share any common factor with $b$, the only possibility for $\delta c_2$ to be holomorphic is $b^2|c_3$. Thus the only possible infinitesimal transformations keeping $\frac{\lambda}{b^2}$, $\frac{\alpha}{b^3}$, $f$ and $g$ invariant are given by (\ref{freetrans}), and
\be
N_r=1+h^0(-K-L).
\ee

In some cases where $L$ has base points, the lower bound (\ref{dh21gineq}) can be improved. For example, when
\be
b|c_3\ ,\ b|c_2\ ,\ b|c_1\label{divcond}
\ee
for any sections $b\in\O(L)$, $c_3\in\O(-K+L)$, $c_2\in\O(-2K)$, $c_1\in\O(-3K-L)$, then we have a larger set of infinitesimal transformations:
\be
\delta b\equiv 0\ ,\ b|\delta c_3.
\ee
In this case, $\delta c_3/b$ is a holomorphic section of line bundle $\O(-K)$, hence
\be
N_r\geq 1+h^0(-K).
\ee

Then the lower bound for $(-\Delta W)$ for general $L$ when (\ref{divcond}) is satisfied is:
\be
\bsp
-\Delta W\geq &h^0(-4K)+h^0(-6K)-h^0(L)-h^0(-4K-2L)-h^0(-3K-L)-h^0(-2K)\\
&-h^0(-K+L)+h^0(-K)+1\label{dh21g2}.
\end{split}
\ee

This case only happens when $h^0(L)=1$ and the single generator $b\in H^0(L)$ of the linear system $|L|$ divides every holomorphic section of $-K+L$, $-2K$ and $-3K-L$, which means that the multiplicative map $s:x\rightarrow bx$ is bijective between the following sets: 
\be
\bsp
H^0(-K)&\xrightarrow{s} H^0(-K+L)\\
H^0(-2K-L)&\xrightarrow{s} H^0(-2K)\\
H^0(-3K-2L)&\xrightarrow{s} H^0(-3K-L).
\end{split}
\ee
This condition is then equivalent to the following relations:
\be
h^0(-K)=h^0(-K+L)\ ,\ h^0(-2K-L)=h^0(-2K)\ ,\ h^0(-3K-2L)=h^0(-3K-L).\label{divcond2}
\ee

\subsection{Minimal tuning of U(1)}

Now we have the following conjecture:

\vspace{0.3cm}

\noindent\textit{Conjecture 1}

The minimal value of $-\Delta h^{d,1}(X)$ and $-\Delta W$ on a given base $B$ when a U(1) is tuned is given by the choice $L=0$.

\vspace{0.3cm}

In the above statement, we have not taken into account the possible bad singularities in the elliptic CY manifold $X$. If the choice $L=0$ leads to (4,6) singularities of $f$ and $g$ over some codimension 1 or 2 base locus, then this choice is not acceptable. Nonetheless, the acceptable values of $-\Delta h^{d,1}(X)$ and $-\Delta W$ are still lower bounded by (\ref{dh21L0}). 

We can construct a sufficient condition which implies the Conjecture 1 using formula (\ref{dh21g}) and (\ref{dh21g2}). We only need to check that for any $L\leq -3K$ that does not satisfy (\ref{divcond2}), the following inequality holds
\be
\bsp
&h^0(-4K)+h^0(-6K)-h^0(L)-h^0(-4K-2L)-h^0(-3K-L)-h^0(-2K)\\
&-h^0(-K+L)+h^0(-K-L)+1\geq h^0(-6K)-h^0(-3K)-h^0(-2K),
\end{split}
\ee
and for any $L\leq -3K$ that satisfies (\ref{divcond2}), the following inequality holds
\be
\bsp
&h^0(-4K)+h^0(-6K)-h^0(L)-h^0(-4K-2L)-h^0(-3K-L)-h^0(-2K)\\
&-h^0(-K+L)+h^0(-K)+1\geq h^0(-6K)-h^0(-3K)-h^0(-2K).
\end{split}
\ee

We have checked that the above statement holds for all the 2D toric bases $B$ with $h^{1,1}(B)\leq 7$ and all the effective line bundles $L\leq -3K$ on them. However, it is hard to rigorously prove the relations between $h^0$ of various line bundles which can be non-ample in general. 

We can make another argument for 2D bases using Green-Schwarz anomaly cancellation conditions in 6D supergravity \cite{Park:2011wv,Morrison-Park}. We prove Conjecture 1 for the cases without non-Abelian gauge groups. The bases $B$ are complex surfaces with no curves with self-intersection $C^2\leq -3$. Mathematically, they are classified as generalized del Pezzo surface with $h^{1,1}(B)\leq 9$ \cite{Derenthal,DerenthalThesis}\footnote{There is another class of 2D F-theory bases with $h^{1,1}(B)=10$ satisfying this condition: elliptic rational surfaces with degenerate elliptic fibers \cite{Persson,Miranda}. But there are always non-Higgsable U(1)s \cite{FiberProduct}, hence we do not consider these cases here.}. Since there is only one gauge group: U(1), the relevant anomaly vectors are $a,b_1\in SO(1,h^{1,1}(B)-1)$ in the 8D anomaly polynomial:
\be
I_8\propto\left(\frac{1}{2}a\mathrm{tr} R^2+2b_1 F^2\right)^2
\ee
Here $R$ is the Ricci 2-form and $F$ is the U(1) field strength 2-form.

In the cases without non-Abelian gauge groups, these anomaly vectors are given by (see (3.18) in \cite{Morrison-Park}):
\be
a=K\ ,\ b_1=-2K+2L.
\ee 

The anomaly cancellation conditions involving $n_i$ U(1) charged hypermultiplets with charge $q_i$ are:
\be
\bsp
a\cdot b_1&=-\frac{1}{6}\sum_i n_i q_i^2\\
b_1\cdot b_1&=\frac{1}{3}\sum_i n_i q_i^4.
\end{split}
\ee

If there are only U(1) charged hypermultiplets with $q_1=\pm 1$ and $q_2=\pm 2$, then we rewrite the above equations as:
\be
\bsp
2K^2-2K\cdot L&=\frac{1}{6}n_1+\frac{2}{3}n_2\\
4K^2-8K\cdot L+4L^2&=\frac{1}{3}n_1+\frac{16}{3}n_2.\label{anomaly2D1}
\end{split}
\ee
We can solve
\be
n_1=12K^2-8K\cdot L-4L^2\ ,\ n_2=L^2-K\cdot L.\label{anomaly12}
\ee
and the total number of charged hypermultiplet
\be
H_{\mathrm{charged}}=n_1+n_2=12K^2-9K\cdot L-3L^2.\label{Hcharged}
\ee

$H_{\mathrm{charged}}$ is related to $-\Delta h^{2,1}(X)$ via the gravitational anomaly constraint:
\be
H_{\mathrm{charged}}+H_{\mathrm{neutral}}-V=273-29T.
\ee
$H_{\mathrm{charged}}$, $V$ and $T$ are the numbers of charged scalar, vector and tensor hypermultiplets. $T=h^{1,1}(B)-1$ does not change in the tuning process. Then from the relation
\be
\Delta H_{\mathrm{charged}}+\Delta H_{\mathrm{neutral}}=\Delta V=1
\ee
after the U(1) is tuned, we get
\be
-\Delta h^{2,1}(X)=-\Delta H_{\mathrm{neutral}}=\Delta H_{\mathrm{charged}}-1.\label{h21charged}
\ee
Here $\Delta H_{\mathrm{charged}}= H_{\mathrm{charged}}$ in (\ref{Hcharged}) since there is no charged matter before the tuning. Nota that we always have $-\Delta h^{2,1}(X)=-\Delta W$ if only a U(1) is tuned, since the degree of vanishing of $\Delta$ is not changed on (-2)-curves.

Now we only need to prove that
\be
-3K\cdot L-L^2\geq 0
\ee
for any $L$ that satisfies $-3K-L\geq 0$.

We use the Zariski decomposition of the effective divisor $L$ \cite{Zariski}:
\be
L=N+P,
\ee
where $P$ is a nef divisor ($P\cdot C\geq 0$ holds for all the curves $C$ on $B$) and $N$ is a linear combination of negative self-intersection curves $N_i$:
\be
N=\sum_i n_i N_i.
\ee
Additionally, $P\cdot N_i=0$ for every $i$ and the intersection matrix $(N_i\cdot N_j)$ is negative definite, such that $N^2\leq 0$. 

Now
\be
-3K\cdot L-L^2=-3K\cdot N-N^2-3K\cdot P-P^2.
\ee
Because $-K$ is a nef divisor for generalized del Pezzo surfaces, $-3K\cdot N\geq 0$. We also have $-N^2\geq 0$ from the negative definiteness of $(N_i\cdot N_j)$. The remaining two terms can be written as
\be
-3K\cdot P-P^2=(-3K-L+N)\cdot P.
\ee
$-3K-L+N$ is effective because $-3K-L$ is effective, and we can conclude $-3K\cdot P-P^2\geq 0$ because $P$ is nef.

We thus proved that
\be
-3K\cdot L-L^2\geq 0.
\ee

If there are charged hypermultiplets with charge $\pm 3$ or higher, we denote the numbers of hypermultiplets with charge $k$ by $n_k$. Then (\ref{anomaly2D1}) is rewritten as:
\be
\bsp
2K^2-2K\cdot L&=\frac{1}{6}n_1+\frac{2}{3}n_2+\frac{1}{6}\sum_{k\geq 3}k^2 n_k\\
4K^2-8K\cdot L+4L^2&=\frac{1}{3}n_1+\frac{16}{3}n_2+\frac{1}{3}\sum_{k\geq 3}k^4 n_k.\label{anomaly2D2}
\end{split}
\ee
Now
\be
\bsp
&n_1=12K^2-8K\cdot L-4L^2-\frac{4}{3}\sum_{k\geq 3}k^2 n_k+\frac{1}{3}\sum_{k\geq 3}k^4 n_k\\ &n_2=L^2-K\cdot L+\frac{1}{12}\sum_{k\geq 3}k^2 n_k-\frac{1}{12}\sum_{k\geq 3}k^4 n_k.
\end{split}
\ee
and the total number of charged hypermultiplets is
\be
H_{\mathrm{charged}}=n_1+n_2+\sum_{k\geq 3}n_k=12K^2-9K\cdot L-3L^2-\frac{5}{4}\sum_{k\geq 3}k^2 n_k+\frac{1}{4}\sum_{k\geq 3}k^4 n_k+\sum_{k\geq 3}n_k.
\ee
Since 
\be
-\frac{5}{4}\sum_{k\geq 3}k^2 n_k+\frac{1}{4}\sum_{k\geq 3}k^4 n_k=\frac{1}{4}\sum_{k\geq 3}k^2(k^2-5)n_k\geq 0,
\ee
the value of $H_{\mathrm{charged}}$ is strictly larger than (\ref{Hcharged}) when there are some hypermultiplets with charge $\pm 3$ or higher.

We hence finished the proof of Conjecture 1 for 2D bases without non-Higgsable clusters.

\subsection{Example: $\mathbb{P}^2$}\label{s:P2}

Now we check the formula (\ref{dh21g}) and (\ref{dh21g2}) for some examples.

The tuning of U(1) on the base $\mathbb{P}^2$ was studied in Morrison and Park's original paper \cite{Morrison-Park}. On $\mathbb{P}^2$, any effective line bundle can be written as $L=nH$, where $H$ is the hyperplane class. The anticanonical class is
\be
-K=3H,
\ee
and the self-intersection of $H$ is $H^2=1$.

The linear system $|nH|$ is the vector space of degree-$n$ homogeneous polynomials in 3 variables (for $n\geq 0$), hence
\be
h^0(nH)=\frac{(n+1)(n+2)}{2}.
\ee
They are all base point free, hence we expect the exact formula (\ref{dh21g}) to hold.

Now the formula of $-\Delta W$ (\ref{dh21L0}) when $L=0$ gives
\be
-\Delta W_{L=0}=h^0(18H)-h^0(9H)-h^0(6H)=107.
\ee
In this case, since there are no (-2)-curves, $\Delta W$ always equals to $\Delta h^{2,1}$.

Since the number of neutral scalar hypermultiplets in the 6d supergravity is
\be
H_{\mathrm{neutral}}=h^{2,1}(X)+1,
\ee
we have
\be
\Delta H_{\mathrm{neutral}}=\Delta h^{2,1}(X)=-107.
\ee

From (\ref{h21charged}), we get the number of charged hypermultiplets that appeared in the tuning process:
\be
H_{\mathrm{charged}}=108,
\ee
which exactly reproduced the result in \cite{Morrison-Park}.

More generally, for $L=nH$, our formula (\ref{dh21g}) gives
\be
\bsp
-\Delta W= &h^0(12H)+h^0(18H)-h^0(nH)-h^0((12-2n)H)-h^0((9-n)H)-h^0(6H)\\
&-h^0((3+n)H)+h^0((3-n)H)+1.\label{dh21gP2}
\end{split}
\ee
When $n\leq 5$, this expression can be reduced to 
\be
-\Delta W=-\Delta h^{2,1}=107+27n-3n^2
\ee
Because of the appearance of a single U(1), $\Delta V=1$, and we know that the number of charged hypermuliplets is
\be
\bsp
H_{\mathrm{charged}}&=\Delta V-\Delta h^{2,1}(X)\\
&=108+27n-3n^2.
\end{split}
\ee

From the anomaly computation (\ref{anomaly12}), we get the numbers $n_1$, $n_2$ of charged hypermultiplets with U(1) charge $\pm1$ and $\pm2$:
\be
(n_1,n_2)=(4(n+3)(9-n),n(n+3)).
\ee
The total number of charged hypermultiplets
\be
H_{\mathrm{charged}}=108+27n-3n^2
\ee
exactly coincides with our formula (\ref{dh21gP2}) when $n\leq 5$.

When $n=6$ or $L=6H$, from (\ref{dh21gP2}) we can compute $-\Delta W=-\Delta h^{2,1}(X)=160$. However, from the anomaly cancellation, the total number of charged hypermultiplets is $H_{\mathrm{charged}}=162$. The discrepancy comes from the fact that another U(1) emerges under this tuning.

Since $L=6H=-2K$, 
\be
c_0\in\mathcal{O}(-4K-2L)=\mathcal{O}(0).
\ee
Hence $c_0$ is a complex number in the Morrison-Park form (\ref{mp}). In this case, we have another rational section apart from (\ref{rationals}):
\be
(x,y,z)=(\frac{1}{4}c_1^2-\frac{2}{3}c_0 c_2,-\frac{1}{8}c_1^3+\frac{1}{2}c_0 c_1 c_2-c_0^2 c_3, c_0^{1/2}).\label{rationals2}
\ee
This rational solution is another Mordell-Weil generator, hence there are two U(1)s under this tuning: $\Delta V=2$. This choice of tuning U(1)$\times$U(1) works for any base.

Now the $-\Delta W$ from (\ref{dh21gP2}) exactly matches the anomaly cancellation condition.

When $n=7$ or 8, $c_0=0$. As discussed at the end of Section~\ref{s:mp}, an additional SU(2) gauge group appears on the (irreducible) curve $c_1=0$. The total gauge groups in these cases are SU(2)$\times$U(1), $\Delta V=4$.

The formula (\ref{dh21gP2}) for $n=7$ and 8 gives $-\Delta W=-\Delta h^{2,1}(X)=146$ and 128 respectively. On the other hand, the number of charged hypermultiplets from anomaly cancellation gives $H_{\mathrm{charged}}=150$ and 132 respectively. Hence our formula (\ref{dh21gP2}) exactly gives the correct number of tuned moduli.

\subsection{Example: $\mathbb{F}_{12}$}\label{s:F12}

\begin{figure}
\centering
\includegraphics[height=6cm]{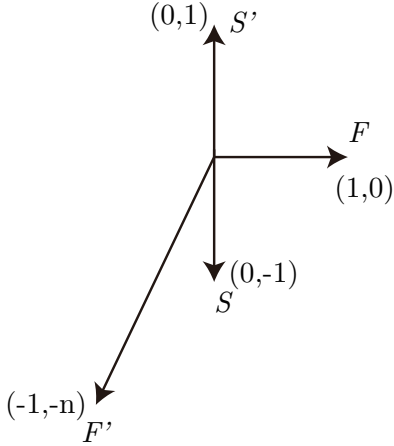}
\caption[E]{\footnotesize The toric fan of the Hirzebruch surface $\mathbb{F}_n$. The toric divisors corresponding to the 1d rays are labelled by $F$, $S$, $F'$ and $S'$. There are two linear relations between them: $F=F'$, $S'=S+nF$.} \label{f:Hirzebruch}
\end{figure}

The Hirzebruch surface $\mathbb{F}_n(n>0)$ is a $\mathbb{P}^1$ bundle over $\mathbb{P}^1$ with twist $n$. The cone of effective divisors on $\mathbb{F}_n$ is generated by a $(-n)$-curve $S$ and the fiber $F$ which is a 0-curve. The easiest way to describe a Hirzebruch surface is to use toric geometry \cite{Fulton,Danilov}. The toric fan of $\mathbb{F}_n$ is shown in Figure~\ref{f:Hirzebruch}. The toric divisors $F$, $S$, $F'$ and $S'$ correspond to rays $(1,0)$, $(0,-1)$, $(-1,-n)$ and $(0,1)$. The linear relations between them are:
\be
F=F'\ ,\ S'=S+nF.
\ee

An arbitrary effective line bundle over $\mathbb{F}_n$ can be parametrized as:
\be
L=aS+bF\ ,\ (a,b\geq 0).
\ee
The anticanonical class is 
\be
-K=F+S+F'+S'=2S+(n+2)F.
\ee
For an arbitrary toric variety with toric divisors $D_i(1<i<n)$ that correspond to 1d rays $v_i\in N=\mathbb{Z}^d$, the holomorphic sections of a line bundle
\be
L=\sum_i a_i D_i
\ee
one-to-one corresponds to the points in the lattice $M=\mathbb{Z}^d$:
\be
S_L=\{p\in\mathbb{Z}^d|\langle p,v_i\rangle\geq -a_i\ ,\ \forall i\}.
\ee

The reason is that if we use the local coordinate description of toric divisors $D_i$: $x_i=0$, then the point $p\in M=\mathbb{Z}^d$ corresponds to the monomial
\be
m_p=\prod_i x_i^{\langle p,v_i\rangle+a_i}.
\ee
in the line bundle $L$. Hence for $m_p$ to be holomorphic, $p$ has to lie in the set $S_L$.

In particular, from the requirement that $f$ and $g$ are holomorphic sections of $\mathcal{O}(-4K)$ and $\mathcal{O}(-6K)$ with no poles, the set of monomials in $f$ and $g$ are given by
\be
\mathcal{F}=\{p\in M=\mathbb{Z}^2|\forall v_i, \langle p,v_i\rangle\geq-4\}\label{toricF}
\ee
and
\be
\mathcal{G}=\{p\in M=\mathbb{Z}^2|\forall v_i, \langle p,v_i\rangle\geq-6\}\label{toricG}
\ee

Now we study F-theory on $X$, where $X$ is a generic elliptic fibration over $\mathbb{F}_n$. For $n\geq 3$, there exists a non-Abelian non-Higgsable gauge group over the $(-n)$-curve $S$. The maximal value for $n$ is 12. When $n>12$, the Weierstrass polynomials $f$ and $g$ vanish to order $(4,6)$ on the curve $S$, which is not allowed in F-theory.

For $\mathbb{F}_{12}$, we denote the divisor $S$ by $s=0$, and $F$ by $t=0$. In the local coordinate patch $SF$, where $s,t$ can vanish and the other local coordinates $x_i$ are set to be 1, $f$ and $g$ can be written as:
\bea
f&=&\sum_{i=4}^8\sum_{j=0}^{12i-40} f_{i,j}t^j s^{i}\\
g&=&\sum_{i=5}^{12}\sum_{j=0}^{12i-60} g_{i,j}t^j s^{i}.
\eea
$f$ and $g$ vanish to order (4,5) on the curve $S$, which gives an $E_8$ gauge group. We can explicitly count
\be
h^0(-4K)=165\ ,\ h^0(-6K)=344.
\ee
The Hodge numbers of the generically fibered elliptic Calabi-Yau threefold $X$ over $\mathbb{F}_{12}$ are $h^{1,1}=11$, $h^{2,1}=491$.

Now we want to tune a U(1) on it. If we take $L=0$, since $h^0(-2K)=51$, $h^0(-3K)=100$, then
\be
-\Delta W_{L=0}=h^0(-6K)-h^0(-3K)-h^0(-2K)=193.
\ee

However, this choice of $L$ is not allowed. The rational section $(x,y,z)=(\lambda,\alpha,1)$ has to obey
\be
\alpha^2=\lambda^3+f\lambda+g.\label{F12eq}
\ee
Now since $\lambda\in\O(-2K)$, $\alpha\in\O(-3K)$, they can be written as
\be
\lambda=\sum_{i=2}^4\lambda_i s^i\ ,\ \alpha=\sum_{i=3}^6\alpha_i s^i.
\ee
Plugging these into (\ref{F12eq}), we can see that the only term of order $s^5$ is the term $g_{5,0}s^5$. Hence this tuning of U(1) requires that $g_{5,0}=0$, which leads to (4,6) singularity on $s=0$.

One way of tuning U(1) on $\mathbb{F}_{12}$ is to take $c_3\in\O(2S')$\label{Johnson-WT}. Since $2S'=2S+24F=-K+10F$, this corresponds to $L=10F$. Now the equation (\ref{F12eq}) becomes
\be
\alpha^2=\lambda^3+f\lambda b^4+gb^6.
\ee
The functions $\lambda\in\O(-2K+2L)$, $\alpha\in\O(-3K+3L)$ can be written as
\be
\lambda=\sum_{i=0}^4\lambda_i s^i\ ,\ \alpha=\sum_{i=0}^6\alpha_i s^i.
\ee
The issue of (4,6) singularity no longer exists.

Now we can compute $-\Delta W$ using our formula (\ref{dh21g}). With $h^0(L)=11$, $h^0(-4K-2L)=h^0(8S+36F)=76$, $h^0(-3K-L)=h^0(6S+32F)=63$, $h^0(-K+L)=h^0(2S+24F)=39$ and $h^0(-K-L)=h^0(4F)=5$, the result is
\be
-\Delta h^{2,1}(X)_{L=10F}= 275.\label{dh21F12}
\ee

On the other hand, $-\Delta H_{\mathrm{neutral}}$ can be computed by anomaly computations \cite{Johnson:2016qar}. Tuning an SU(2) gauge group on the curve $c_3=0$ with self-intersection $n$ and genus $g$ results in charged matter $(6n+16-16g)\mathbf{2}+g\cdot\mathbf{3}$. This SU(2) can be Higgsed to U(1), and the resulting charged matter is $(6n+16-16g)(\mathbf{\pm 1})+(g-1)(\mathbf{\pm 2})$. The change in $H_{\mathrm{neutral}}$ and $h^{2,1}(X)$ is then
\be
\Delta H_{\mathrm{neutral}}=\Delta h^{2,1}(X)=-12n+30(g-1)+1.\label{dh21ng}
\ee

In our case, the curve $c_3=0$ which is $2S'$ has self-intersection $n=48$ and genus $g=11$. This formula exactly gives
\be
\Delta H_{\mathrm{neutral}}=\Delta h^{2,1}(X)=-275,
\ee
perfectly matches (\ref{dh21F12}), as there are no (-2)-curves and we always have $\Delta W=\Delta h^{2,1}$. 

If we set $c_3=0$ to be $3S'$, which corresponds to the choice $L=S+22F$, then our formula (\ref{dh21g}) gives
\be
-\Delta W_{L=10F}=305.
\ee
The curve $3S'$ has self-intersection $n=108$ and genus $g=34$, and (\ref{dh21ng}) gives
\be
\Delta H_{\mathrm{neutral}}=\Delta h^{2,1}(X)=-305,
\ee
exactly matches our formula.

When $c_3=0$ is an element of line bundle $4S'$, or $L=2S+34F$, $c_0=0$ because $-4K-2L$ is not effective. $c_1$ vanishes to order 6 on the curve $s=0$. The discriminant $\Delta$ takes the form
\be
\Delta=\frac{27}{16}b^4 c_1^4+b^2 c_1^2 c_2^3-\frac{9}{2}b^2 c_1^3 c_2 c_3-c_1^2 c_2^2 c_3^2+4 c_1^3 c_3^3,
\ee
which vanishes to degree 12 on the curve $s=0$. Hence this is not valid in F-theory.

\subsection{Example: $\mathbb{F}_3$}

Now we investigate the case $B=\mathbb{F}_3$, whose anticanonical class is $-K=2S+5F$.

First we look at the case $L=0$, which is supposed to be the minimal tuning of U(1). In this case, the formula (\ref{dh21L0}) gives $-\Delta W\equiv-\Delta h^{2,1}(X)=99$.

Denote the (-3)-curve $S$ on $\mathbb{F}_3$ by $s=0$ and the 0-curve $F$ by $t=0$, then $c_3\in\mc{O}(-K)$ can be written as:
\be
c_3=s[(c_{3,0}+c_{3,1}t+c_{3,2}t^2)+\mc{O}(s)],
\ee
where $c_{i,j}$ are numerical coefficents. One can see that the curve $c_3=0$ is a reducible curve containing the (-3)-curve $t=0$. The other component $C=-K-S=S+5F$ is a rational curve with self-intersection 7 which intersects $t=0$ at 2 points. 

The functions $c_0,c_1,c_2$ have the form
\be
\bsp
c_0&=s^2[(c_{0,0}+c_{0,1}t+c_{0,2}t^2)+\mc{O}(s)]\\
c_1&=s[c_{1,0}+\mc{O}(s)]\\
c_2&=s[(c_{2,0}+c_{2,1}t)+\mc{O}(s)].
\end{split}
\ee

If we set $b=0$, then we have
\be
\bsp
f&=c_1 c_3-\fracs{1}{3}c_2^2=As^2+O(s^3)\\
g&=c_0 c_3^2+\fracs{2}{27}c_2^3-\fracs{1}{3}c_1 c_2 c_3=B s^3+O(s^4).
\end{split}
\ee
$A=f_0+f_1 s+f_2 s^2$ and $B=g_0+g_1 s+g_2 s^2+g_3 s^3$ are two polynomials with general coefficients. In this case, the degree of vanishing of $(f,g)$ on $t=0$ is $(2,3)$, and the gauge group is determined by the reducibility of cubic polynomial $h(\psi)=\psi^3+A\psi+B$ \cite{GrassiMorrison}. If $h(\psi)$ has three components, then the gauge group is SO(8); if $h(\psi)$ has two components, then the gauge group is SO(7); if $h(\psi)$ is irreducible, then the gauge group is $G_2$.

For generic coefficients in $A$ and $B$, we can write
\be
\psi^3+(f_0+f_1 s+f_2 s^2)\psi+(g_0+g_1 s+g_2 s^2+g_3 s^3)=(\psi+a_0+a_1 s)(\psi^2+(a_2+a_3 s)\psi+a_4+a_5 s+a_6 s^2),
\ee
since there are exactly seven variables $a_0\sim a_6$ on the r.h.s. Hence in the phase $b=0$, the SU(3) gauge group on $S$ is enhanced to $SO(7)$, and this $SO(7)$ intersects the curve $C$ that carries gauge group SU(2) at two points. This case is similar to the case of -2/-3/-2 non-Higgsable cluster, and the charged matter contents under gauge group SO(7)$\times$SU(2) at the intersection points are $2\cdot(\mathbf{8}_s,\frac{1}{2}\mathbf{2})$ \cite{clusters, Johnson:2016qar}.

Tuning SU(2) on a rational curve with self-intersection 7 leads to matter contents $58\cdot \mathbf{2}$ in total, hence there are 50 copies of charged hypermultiplets in the fundamental representations on the curve $C$ and does not intersect $S$ with SO(7). After the gauge group SO(7)$\times$SU(2) is Higgsed to SU(3)$\times$U(1), they are the reminiscent charged matters under U(1) with charges $\pm1$. Hence there are 100 charged hypermultiplets in the minimal tuning of U(1), which exactly matches the calculation via (\ref{dh21L0}).

Now we investigate the cases in which $c_3\in\mc{O}(kS')$, so that the curve $c_3=0$ does not intersect the (-3)-curve $S$ with SU(3) gauge group on it. We list the computed $-\Delta W=-\Delta h^{2,1}(X)$ from (\ref{dh21g}) and the number of charged hypermultipets in Table~\ref{t:F3nS}.

\begin{table}
\centering
\begin{tabular}{|c|c|c|c|c|c|}
\hline
L & -K+L & $-\Delta W=-\Delta h^{2,1}(X)$ from (\ref{dh21g}) & $n$ & $g$ & $H_{\textrm{charged}}=12n-30(g-1)$\\
\hline
$F$ & $2S'$ & 113 & 12 & 2& 114\\
$S+4F$ & $3S'$ & 143 & 27 & 7 & 144\\
$2S+7F$ & $4S'$ & 155 & 48 &15 & 156\\
$3S+10F$ & $5S'$ & 148 & 75 & 26 & 150\\
\hline
\end{tabular}
\caption{Tuning U(1) on $\mathbb{F}_3$ with $L=F+kS'$. We listed the computed values of $-\Delta W=-\Delta h^{2,1}(X)$ from (\ref{dh21g}) and number of charged hypermultipets. $n$ and $g$ denotes the self-intersection number and $g$ of the curve $c_3=0$. }\label{t:F3nS}
\end{table}

For the cases $L=F,\ S+4F,\ 2S+7F$, there is only one gauge group U(1), and $-\Delta W$ from (\ref{dh21g}) explicitly matches the expected number of charged hypermultiplets.

For $L=3S+10F$, $c_0\in\mathcal{O}(-4K-2L)=\mathcal{O}(2S)$ is a perfect square. Hence the other rational section (\ref{rationals2}) appears, and there are two U(1)s in the supergravity theory. This indeed matches the numbers in Table~\ref{t:F3nS}.

Another type of $L$ is $L=kF$, with $k>1$. First let us study the case $L=2F$. In this case, even when $b\neq 0$, the gauge group on the (-3)-curve is already enhanced to $G_2$. This enhancement to $G_2$ leads to charged matter $1\cdot\mathbf{7}$ in the fundamental representation of $G_2$, and the change in vector hypermultiplet $\Delta V=14-8=6$. However, this charged matter $1\cdot\mathbf{7}$ only contributes 6 to $\Delta H_{\textrm{charged}}$, because one component of the $\mathbf{7}$ is neutral under the Cartan subgroup of $G_2$ \cite{GrassiMorrison}. This fact can be seen from the branching rule of $G_2\rightarrow SU(3)$:
\be
\mathbf{7}\rightarrow\mathbf{3}+\bar{\mathbf{3}}+\mathbf{1}.
\ee
The singlet component $\mathbf{1}$ is neutral under the Cartan subgroup of SU(3), hence it is neutral under the Cartan subgroup of $G_2$ because the rank of $G_2$ and SU(3) are the same. 

Hence the enhancement of $G_2$ on a (-3)-curve does not change $h^{2,1}(X)$. If we set $b=0$ and unHiggs the U(1) to SU(2), since the curve $c_3=0$ has self-intersection $n=16$ and genus $g=3$, the total matter content on $c_3=0$ is $64\cdot\mathbf{2}+3\cdot\mathbf{3}$. For the matter at the intersection point of $c_3=0$ and the (-3)-curve, the anomaly cancellation conditions are the same as the non-Higgsable cluster -2/-3. Hence the total charged matter contents in the phase $b=0$ under gauge groups $G_2\times$ SU(2) is $(\mathbf{7}+\mathbf{1},\frac{1}{2}\mathbf{2})+60\cdot(1,\mathbf{2})+3\cdot(1,\mathbf{3})$. After $G_2\times$ SU(2) is Higgsed to $G_2\times$ U(1), the matter contents are $(\mathbf{7}+\mathbf{1},\frac{1}{2}(\pm\mathbf{1}))+60\cdot(1,\pm\mathbf{1})+2\cdot(1,\pm\mathbf{2})$. In total, $\Delta H_{\mathrm{charged}}=132$, hence the change in $h^{2,1}(X)$ from the non-Higgsable phase is given by:
\be
-\Delta h^{2,1}(X)=-\Delta H_{\mathrm{neutral}}=\Delta H_{\mathrm{charged}}-\Delta V=132-6-1=125.
\ee
This exactly matches the result computed with the formula (\ref{dh21g}).

Similarly, for the case $L=3F$, the gauge group on the (-3)-curve is enhanced to SO(7). $H_{\mathrm{charged}}$ on the curve $c_3=0$ is 150, and the change in $h^{2,1}(X)$ from the non-Higgsable phase:
\be
-\Delta h^{2,1}(X)=-\Delta H_{\mathrm{neutral}}=\Delta H_{\mathrm{charged}}-\Delta V=150-21+8-1=136,
\ee
exactly matches (\ref{dh21g}).

For the case $L=kF$, $(k\geq 4)$, the gauge group on the (-3)-curve is enhanced to SO(8). We will not discuss the anomaly computations in detail.

\section{Constraints on bases with a non-Higgsable U(1)}\label{s:constraints}

The existence of a non-Higgsable U(1) on a base implies that for generic $f\in\O(-4K)$ and $g\in\O(-6K)$, the Weierstrass model has an additional rational section and can be written in the Morrison-Park form (\ref{mp})\footnote{Again, we do not consider the cases where the Weierstrass model cannot be written in Calabi-Yau Morrison-Park form \cite{Klevers-WT,TallSection}. In these exotic cases, U(1) charged matter with charge $q\geq 3$ appears. We assume that for the cases of non-Higgsable U(1)s without matter, the Weierstrass model can always be written in the Morrison-Park form. This includes the case of multiple U(1)s, which corresponds to a further refinement of the coefficients in the Morrison-Park form.}. Assuming Conjecture 1 holds, this condition means that the minimal value of $-\Delta W$ in the formula (\ref{dh21L0}) is non-positive:
\be
-\Delta W_{L=0}=h^0(-6K)-h^0(-3K)-h^0(-2K)\leq 0\label{constraintdW}
\ee
This inequality imposes stringent constraint on the Newton polytopes $A_n$ of $\O(-nK)$. For toric bases, they are the polytopes defined by the set of lattice points:
\be
A_n=\{p\in\mathbb{Z}^d|\langle p,v_i\rangle\geq -n\ ,\ \forall i\}.
\ee
Note that $A_4$ and $A_6$ corresponds to $\mathcal{F}$ and $\mathcal{G}$ defined in (\ref{toricF}) and (\ref{toricG}) respectively.

The notion of Newton polytopes can be generalized to arbitrary bases. For a point $p=(x_1,x_2,\dots,x_d)$ in the Newton polytope $A_n$, it corresponds to a monomial $m_p=\alpha^n\prod_{i=1}^d \beta_i^{x_i}$, where $\alpha$ and $\beta_i$ are some non-zero functions. Hence the product of two monomials is mapped to the vector sum of two points in $\mathbb{Z}^d$. For example, the expression of $f$ and $g$ for a semi-toric base with non-Higgsable U(1)s in Section~\ref{s:2D} can be written as (\ref{fggdp}). In that case, we can assign $\alpha=\tilde{\eta}$, $\beta_1=\eta\tilde{\eta}^{-1}$, and the Newton polytopes for $f$ and $g$ are one-dimensional. Note that the origin in the $\mathbb{Z}^d$ can be shifted.

We use the notation $nP$ to denote the set of lattice points in the resized polytope enlarged by a factor $n$:
\be
nP=\{p|p=\sum_{i=1}^n p_i\ ,\ \forall p_i\in P\}.
\ee
The lattice points in $nP$ correspond to the monomials $m=\prod_{i=1}^n m_i$, where $m_i\in P$. $|A|$ denotes the number of lattice points in the Newton polytope $A$. 

We denote the dimension of the Newton polytope $A_n$ of $\O(-nK)$ by $d_{An}$. Clearly the dimension of any other $A_n(n<6)$ is equal to or smaller than $d_{A6}$. We want to prove that $d_{A6}\leq 1$ if (\ref{constraintdW}) holds.

Now if the dimension of the set $A_3$, $d_{A3}$, is higher than 1, then we always have $|2A_3|\geq 2|A_3|$. This can be argued by classifying the points $p\in A_3, p\neq \vec{0}$ with the ray $r=p\vec{0}$ that connects $p$ to the origin. If there are $k$ points on such a ray $r$ (not including the origin), then in the polytope $2A_3$, we have at least $2k$ points on the same ray $r$ (not including the origin). Taking account of the origin in $2A_3$, now we already have at least $2|A_3|-1$ points in $2A_3$. Furthermore, if $d_{A3}>1$, we can pick two points $p_1\neq\vec{0}$ and $p_2\neq\vec{0}$ on the boundary of $A_3$, such that there is no point $p\in A_3$ that lies on the line segment $p_1 p_2$. Then $p_1+p_2$ gives another point in $2A_3$ that has not been counted yet. If $A_3$ does not include the origin $\vec{0}$, we can shift all the points in the polytope by a constant vector and transform one of the lattice point to $\vec{0}$, then the argument still works.

Similarly, if the dimension of the set $A_2$, $d_{A2}$, is higher than 1, then we always have $|3A_2|\geq 3|A_2|$. This is similarly argued by classifying the points $p\in A_2, p\neq \vec{0}$ with the ray $r=p\vec{0}$. If there are $k$ points on such a ray $r$ (not including the origin), then in the polytope $3A_2$, we have at least $3k$ points on the same ray $r$ (not including the origin). Taking account of the origin in $3A_2$, now we already have at least $3|A_2|-2$ points in $3A_2$. Furthermore, if $d_{A3}>1$, we can pick two points $p_1\neq\vec{0}$ and $p_2\neq\vec{0}$ on the boundary of $A_2$, such that there is no point $p\in A_2$ lying on the line segment $p_1 p_2$. Then $2p_1+p_2$ and $p_1+2p_2$ are two additional points in $3A_2$. 

Note that regardless of $d_{A2}$, $d_{A3}$ and $d_{A6}$, we always have
\be
|2A_3|\geq 2|A_3|-1\ ,\ |3A_2|\geq 3|A_2|-2\label{universalA23}
\ee

Then since $2A_3\subseteq A_6$ and $3A_2\subseteq A_6$,
\be
\bsp
-\Delta W_{L=0}&=|A_6|-|A_3|-|A_2|\\
&\geq |2A_3|-|A_3|-|A_2|\\
&\geq |A_3|-|A_2|-1
\end{split}
\ee
and
\be
\bsp
-\Delta W_{L=0}&=|A_6|-|A_3|-|A_2|\\
&\geq |3A_2|-|A_3|-|A_2|\\
&\geq 2|A_2|-|A_3|-2.
\end{split}
\ee

If $-\Delta W_{L=0}\leq 0$ so that non-Higgsable U(1)s appear, we have constraints on $|A_3|$ and $|A_2|$:
\be
|A_3|-|A_2|-1\leq 0\ ,\ 2|A_2|-|A_3|-2\leq 0.
\ee
The only possible values for $(|A_2|,|A_3|)$ satisfying these inequalities are $(1,1)$, $(1,2)$, $(2,2)$, $(2,3)$ and $(3,4)$. We list the possible values of $|A_6|$ for each of these cases in Table~\ref{t:nHU1A}. 

\begin{table}
\centering
\begin{tabular}{|c|c|c|c|c|c|}
\hline
$|A_2|$&1&1&2&2&3\\
\hline
$|A_3|$&1&2&2&3&4\\
\hline
$|A_4|$&1,2&1,2,3&3&3,4&5\\
\hline
$|A_6|$&2&3&4&5&7\\
\hline
\end{tabular}
\caption{The possible values of the number of lattice points in the Newton polytope $A_n$ of $\O(-nK)$, when non-Higgsable U(1)s exist.}\label{t:nHU1A}
\end{table}

Now we argue that the dimension of Newton polytopes $A_2$, $A_3$ and $A_6$ cannot be higher than 1. The statement is trivial for $|A_6|=2$ because these polytopes have at most 2 points. If $|A_6|=3$, since $|A_6|=2|A_3|-1$, $|2A_3|\geq 2|A_3|-1$ and $|2A_3|\leq|A_6|$ (see (\ref{universalA23})), we know that $|2A_3|=2|A_3|-1$ and $A_6$ coincides with the polytope $2A_3$. Because $A_3$ is one-dimensional, we conclude that $d_{A6}=d_{A3}=1$. If $|A_6|=4$, since $|A_6|=|3A_2|=3|A_2|-2$, $A_6$ coincides with $3A_2$ and $d_{A6}=d_{A2}=1$. If $|A_6|=5$, similarly because $|A_6|=|2A_3|=2|A_3|-1$, $A_6$ coincides with the polytope $2A_3$. We have argued that if $d_{A3}>1$, then $|2A_3|\geq 2|A_3|$. This does not happen here, hence the polytopes $A_3$ and $A_6$ are one-dimensional. When $|A_6|=7$, similarly $|A_6|=|2A_3|=2|A_3|-1$, then $A_6$ coincides with the polytope $2A_3$ and they are both one-dimensional.

We also list the possible values of $|A_4|$ in Table~\ref{t:nHU1A}. They can be computed after the Newton polytopes $A_n$ are linearly transformed to integral points on line segments $[na,nb]$, $a,b\in\mathbb{Q},a<b$. This SL$(d,\mathbb{Q})$ linear transformation is always possible because the Newton polytopes $A_n$ are one-dimensional. Then for each $|A_6|$, we try to construct the pairs $(a,b)$ which give all the possible value of $|A_4|$ with the correct values of $|A_2|$ and $|A_3|$. For $|A_6|=2$, we can take $(a,b)=(0,\frac{1}{4})$ to get $|A_4|=2$ and $(a,b)=(0,\frac{1}{6})$ to get $|A_4|=1$. For $|A_6|=3$, we can take $(a,b)=(-\frac{1}{4},\frac{1}{4})$ to get $|A_4|=3$, $(a,b)=(0,\frac{1}{3})$ to get $|A_4|=2$ and $(a,b)=(\frac{1}{3},\frac{2}{3})$ to get $|A_4|=1$. For $|A_6|=4$, we can only take $(a,b)=(0,\frac{1}{2})$ to get $|A_4|=3$. For $|A_6|=5$, we can take $(a,b)=(0,\frac{3}{4})$ to get $|A_4|=4$ and $(a,b)=(0,\frac{2}{3})$ to get $|A_4|=3$. For $|A_6|=7$, we can only take $(a,b)=(0,1)$ to get $|A_4|=5$.

Hence we have the bounds on the number of monomials in $f$ and $g$:
\be
|A_4|\leq 5\ ,\ |A_6|\leq 7\label{FGbound}
\ee

This shows that the number of monomials in $A_4$ and $A_6$ are very small when non-Higgsable U(1)s exist. Recall the formula for $h^{d,1}$ (\ref{hd1}), we expect the number of complex structure moduli of the elliptic Calabi-Yau manifold over this base to be small. In the 4D F-theory context, it means that the number of flux vacua is small\cite{Douglas:2006es,Denef-F-theory,Ashok-Douglas,Denef-Douglas,Bousso-Polchinski,Braun-Watari1,Watari,MostFluxVacua}.

Furthermore, the one-dimensional feature of polytopes $A_n$ excludes the existence of non-Higgsable U(1)s on any smooth compact toric bases.

This statement can be argued as follows.

\begin{figure}
\centering
\includegraphics[height=6cm]{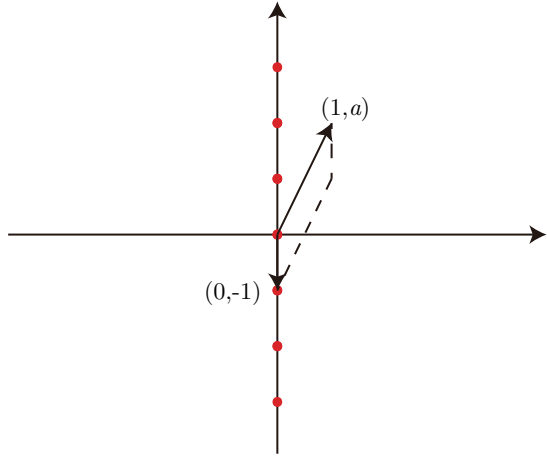}
\caption[E]{\footnotesize A fictional configuration of 2D toric base with monomials in $g$ aligned along the $y$-axis. Suppose that there is no ray (1,0), then this is the only possibility up to a linear transformation.} \label{f:2Dalign}
\end{figure}

For the case of 2D toric bases, we perform an SL$(2,\mathbb{Z})$ transformation on the 2D toric fan, such that the monomials in $g$ align along the $y$-axis. Now, we observe that there always exists a ray (1,0) in the fan. Otherwise, the only possible 2D cone for a smooth 2D toric base near the positive $x$-axis consists of a ray $(1,a)$ and a ray $(-b,-ab-1)$, where $a\geq 1,b\geq 0$, see Figure~\ref{f:2Dalign}.

If this is the case, suppose that (0,1) is in the Newton polytope $A_6$ for $g$, which means that there is no ray $(x,y)$ in the fan with $y<-6$. Then from the structure of the fan, we can see that $(-1,1)$ is also in the polytope $A_6$. This is because the other rays $(x,y)$ with $x>0$ all satisfy $y>ax$, hence they cannot satisfy $(x,y)\cdot(-1,1)=-x+y<-6$. Then we conclude that $A_6$ is not a one-dimensional polytope, which contradicts our assumption. Hence $(0,1)\notin A_6$. Similarly, we can exclude the existence of all points $(0,y>0)$ in $A_6$. However, this implies that $g$ vanishes to order 6 on the divisor corresponding to the ray $(-b,-ab-1)$. Because $A_4\subset A_6$ for toric bases, this excludes all the points $(0,y>0)$ in $A_4$ as well. Hence $(f,g)$ vanishes to order $(4,6)$ on the divisor corresponding to the ray $(-b,-ab-1)$, which is not allowed.

Hence we conclude that a ray (1,0) has to exist in the fan. However, because $A_4$ and $A_6$ aligns along the $y$-axis, it means that $(f,g)$ vanishes to order $(4,6)$ on the divisor corresponding to the ray (1,0). So this is not allowed, either.

Similar arguments can be applied to higher dimensional cases. For any smooth compact toric bases, there has to be a ray perpendicular to the line on which $f$ and $g$ aligns, but this will lead to (4,6) singularity on such a ray. If this ray does not exist, then there will be (4,6) singularities over codimension-two locus on the base that cannot be resolved by blowing up this locus.

We elaborate this statement for the case of toric threefold bases. We assume that the Newton polytopes $A_4$ and $A_6$ lie on the $z$-axis, and the fan of the toric base has no ray on the plane $z=0$. Because the base is compact, there exists a 2D cone $v_1 v_2$ in the fan such that $v_{1z}>0$ and $v_{2z}<0$ ($v_{1z}$ and $v_{2z}$ are the $z$-components of $v_1$ and $v_2$ respectively). Now we can analyze the degree of vanishing of $(f,g)$ on this codimension-two locus $v_1 v_2$ for each of the cases in Table~\ref{t:nHU1A}. For example, if $|A_4|=5$, $|A_6|=7$, such that the points in $A_4$ are $(0,0,-2)\sim(0,0,2)$ and the points in $A_6$ are $(0,0,-3)\sim(0,0,3)$, then we have constraints on $v_{1z}$ and $v_{2z}$: $2\geq v_{1z}\geq 1$, $-1\geq v_{2z}\geq -2$. Now if $v_{1z}=-v_{2z}=1$ or $v_{1z}=-v_{2z}=2$, then it is easy to see that the degree of vanishing of $(f,g)$ on this curve $v_1 v_2$ is $(8,12)$. We cannot resolve this by blowing up the curve $v_1 v_2$, because the ray of the exceptional divisor will lie on the plane $z=0$, which contradicts our assumptions. If $v_{1z}=2$, $v_{2z}=-1$ or $v_{1z}=1$, $v_{2z}=-2$, then the degree of vanishing of $(f,g)$ on this curve $v_1 v_2$ is $(6,9)$. If we try to resolve this by blowing up the curve $v_1 v_2$, then the exceptional divisor corresponds to a ray $v_3=v_1+v_2$ with $v_{3z}=\pm 1$. Then the $(4,6)$ singularity issues remains on the curve $v_2 v_3$ or $v_1 v_3$. Hence we cannot construct a good toric threefold base with such Newton polytopes $A_4$ and $A_6$. Similarly we can explicitly apply this argument to all the other possible configurations of $A_4$ and $A_6$, showing that it is impossible to construct a toric threefold base with non-Higgsable U(1)s and without any codimension-one or codimension-two (4,6) singularities. This argument is independent of the dimension of the toric base, either.

This 1D feature of Weierstrass polynomials suggests that the bases with non-Higgsable U(1) always have the structure of a fibration over $\mathbb{P}^1$. 

We have the following conjecture:

\noindent\textit{Conjecture 2:}
Any $n$-dimensional base with non-Higgsable U(1) can be written as a resolution of a Calabi-Yau $(d-1)$-fold fibration over $\mathbb{P}^1$. The generic fiber is a smooth Calabi-Yau $(d-1)$-fold\footnote{The author thanks Daniel Park and David Morrison for useful discussions involved in the following explanations of this conjecture}.

The alignment of $f\in\mc{O}(-4K_B)$ and $g\in\mc{O}(-6K_B)$ on a line suggests that the base $B$ is either a fibration of $F=-K_B$ or a blow up of such a $-K_B$ fibration. One can understand this from an analogous toric setup, where $B$ is a Hirzebruch surface $\mathbb{F}_n$ which is a $\mathbb{P}^1$ bundle over $\mathbb{P}^1$ (see the geometric description at the beginning of Section~\ref{s:F12}). Because of the $\mathbb{P}^1$ fibration structure, one can see that any line bundle $\mc{O}(nF)$ on $\mathbb{F}_n$ has an one-dimensional Newton polytope. More generally, if we blow up $\mathbb{F}_n$ and the 0-curve $F$ on $\mathbb{F}_n$ remains, the line bundle $\mc{O}(nF)$ still has an one-dimensional Newton polytope.

Now return to our case $F=-K_B$. From the adjunction formula, we can see that the canonical class of the fiber $F=-K_B$ vanishes:
\be
K_F=(K_B+F)|_F=0,
\ee
hence the fiber $F$ is Calabi-Yau.

Additionally, the fiber $F$ should not self-intersect, hence we have
\be
F\cdot F=0.
\ee

The elliptic Calabi-Yau $(d+1)$-fold $X$ over this base $B$ can be thought as resolution of the fiber product space, constructed below.

Take a rational elliptic surface $A$ with section
\be
\pi_A: A\rightarrow\mathbb{P}^1
\ee
and a Calabi-Yau $(d-1)$-fold fibration $B$ with section
\be
\pi_B: B\rightarrow\mathbb{P}^1
\ee

Then the elliptic Calabi-Yau $(d+1)$-fold $X$ is the resolution of the fiber product
\be
\tilde{X}=A\times_{\mathbb{P}^1} B=\{(u,v)\in A\times B|\pi_A(u)=\pi_B(v)\}.
\ee
In the case of $d=2$, this is the generalized Schoen construction of fiber products of rational elliptic surfaces \cite{FiberProduct}.

In the case of $d=3$, the base $B$ is a resolution of a K3 or $T^4$ fibration over $\mathbb{P}^1$.

Now we have an alternative interpretation of the relation between the fibration structure of $B$ and the 1D feature of Weierstrass polynomials from the pullback of $f$ and $g$ over the elliptic surface $A$ \cite{FiberProduct}:
\be
\bsp
f_{X'}&=\pi^\star_B(f_{\bar{A}})\\
g_{X'}&=\pi^\star_B(g_{\bar{A}}).
\end{split}
\ee
Here $X'$ is the Weierstrass model over the base $B$, which is possibly singular. $\bar{A}$ is the possibly singular Weierstrass model over $\mathbb{P}^1$. $X'$ and $\bar{A}$ are related by
\be
X'=\bar{A}\times_{\mathbb{P}^1} B=\{(\bar{u},v)\in \bar{A}\times B|\pi_{\bar{A}}(\bar{u})=\pi_B(v)\}.
\ee
Hence the 1D property of Weierstrass polynomials $f_{X'}$ and $g_{X'}$ over $B$ is inherited from the 1D property of $f_{\bar{A}}$ and $g_{\bar{A}}$.

\section{Semi-toric generalized Schoen constructions}\label{s:2D}

\begin{figure}
\centering
\includegraphics[height=3cm]{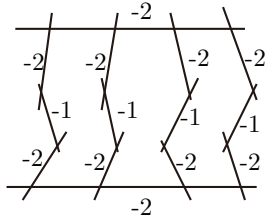}
\caption[E]{\footnotesize A description of generalized $dP_9$ with the negative curves on it. Note that there are two copies of (-2)-curve clusters that correspond to degenerate Kodaira fibers of type $I_0^*$.} \label{f:gdP9}
\end{figure}

As the first example with a non-Higgsable U(1), we choose the base to be the semi-toric generalized $dP_9$ constructed in  \cite{Martini-WT}. We call it $gdP_{9st}$, and we have $h^{1,1}(gdP_{9st})=10$, $T=9$. The set of negative curves on $gdP_{9st}$ is shown in Figure~\ref{f:gdP9}. 

Consider the general elliptic CY3 $X$ over $gdP_{9st}$. In \cite{Martini-WT}, it is computed that $h^{1,1}(X)=h^{2,1}(X)=19$. There is a $U(1)^8$ abelian gauge group in the 6D low-energy effective theory, which explains the rank of the gauge group ($h^{1,1}(X)=h^{1,1}(gdP_{9st})+$rk$(G)+1=19$, rk$(G)=8$). Also, we have the anomaly cancellation in 6D:
\be
273-29T=H-V=h^{2,1}(X)+1-V.
\ee
When $T=9$, $h^{2,1}(X)=19$, we get the correct number of vector multiplets $V=8$.

\begin{figure}
\centering
\includegraphics[height=3cm]{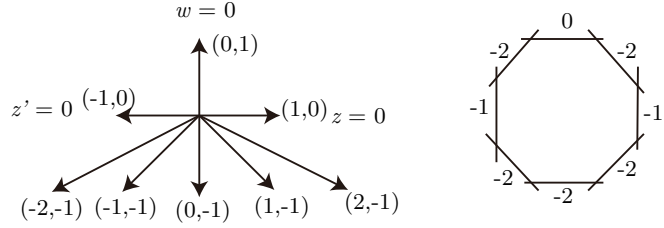}
\caption[E]{\footnotesize A toric generalized $dP_5$, with toric cyclic diagram $(0,-2,-1,-2,-2,-2,-1,-2)$, and the 2D toric rays shown on the left.} \label{f:2Dtoric}
\end{figure}

To write down the set of monomials in $f$ and $g$ for a general elliptic fibration over $gdP_{9st}$, we start with the 2D toric base with toric diagram in Figure~\ref{f:2Dtoric}.

\begin{figure}
\centering
\includegraphics[height=4cm]{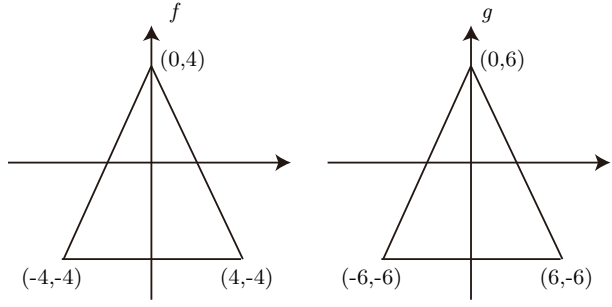}
\caption[E]{\footnotesize The set of monomials in $f$ and $g$, for the toric generalized $dP_5$ described in \ref{f:2Dtoric}} \label{f:2Dtoricfg}
\end{figure}

Using (\ref{toricF}) and (\ref{toricG}), we can compute the Newton polytope of Weierstrass polynomials $f$ and $g$, shown in Figure~\ref{f:2Dtoricfg}.

We then write down the general expression of $f$ and $g$ in the patch $(z',w)$ and $(z,w)$:
\be
\bsp
f&=\sum_{i=0}^8 f_{0,i}z^i z'^{8-i}+w\sum_{i=1}^7 f_{1,i}z^i z'^{8-i}+w^2\sum_{i=1}^7 f_{2,i}z^i z'^{8-i}+w^3\sum_{i=2}^6 f_{3,i}z^i z'^{8-i}\\
&+w^4\sum_{i=2}^6 f_{4,i}z^i z'^{8-i}+w^5\sum_{i=3}^5 f_{5,i}z^i z'^{8-i}+w^6\sum_{i=3}^5 f_{6,i}z^i z'^{8-i}+w^7 f_{7,4}z^4 z'^4+w^8 f_{8,4}z^4 z'^4.
\end{split}
\ee

\be
\bsp
g&=\sum_{i=0}^{12} g_{0,i}z^i z'^{12-i}+w\sum_{i=1}^{11} g_{1,i}z^i z'^{12-i}+w^2\sum_{i=1}^{11} g_{2,i}z^i z'^{12-i}+w^3\sum_{i=2}^{10} g_{3,i}z^i z'^{12-i}\\&+w^4\sum_{i=2}^{10} g_{4,i}z^i z'^{12-i}
+w^5\sum_{i=3}^9 g_{5,i}z^i z'^{12-i}+w^6\sum_{i=3}^9 g_{6,i}z^i z'^{12-i}+w^7\sum_{i=4}^8 g_{7,i}z^i z'^{12-i}\\&+w^8\sum_{i=4}^8 g_{8,i}z^i z'^{12-i}+w^9\sum_{i=5}^7 g_{9,i}z^i z'^{12-i}+w^{10}\sum_{i=5}^7 g_{10,i}z^i z'^{12-i}+w^{11} g_{11,6} z^6 z'^6\\&+w^{12}g_{12,6} z^6 z'^6
\end{split}
\ee

In the above expressions, we have set all the local coordinates apart from $z',z$ and $w$ to be 1. From now on, we generally work in the patch $(z,w)$, so we set $z'=1$. To recover the dependence on $z'$, one just needs to multiply the factor $z'^{8-i}$ to each $z^i$ term in $f$, and multiply $z'^{12-i}$ to each $z^i$ term in $g$. By the way, we will change the definition of coefficients $f_{ij}$ and $g_{ij}$ very often, they only mean general random complex numbers, for a generic fibration.

Now we blow up a point $z=1$, $w=0$, which is a generic point on the divisor $w=0$. After the blow up, we assign new coordinates $z_1,w_1$ and $\xi_1$:
\be
z-1=(z_1-1)\xi_1\ ,\ w=w_1\xi_1\label{blp1}
\ee
The divisor $w=0$ becomes a (-1)-curve $w_1=0$. $\xi_1=0$ is the exceptional divisor of this blow-up, and $z_1=1$ is a new (-1)-curve. We plug (\ref{blp1}) into the expression of $f$ and $g$. Note that all the terms with $\xi_1^m$ in $f$ vanish for $m<4$, similarly all the terms with $\xi_1^m$ in $g$ vanish for $m<6$. This impose constraints on the coefficients $f_{ij}$ and $g_{ij}$. Finally, we divide $f$ by $\xi_1^4$, and $g$ by $\xi_1^6$ after the process is done. The resulting $f$ can be written as:
\be
\bsp
f&=(z_1-1)^4\sum_{i=0}^4 f_{0,i}z^i+(z_1-1)^3 w_1\sum_{i=1}^4 f_{1,i}z^i+(z_1-1)^2 w_1^2\sum_{i=2}^5 f_{2,i}z^i+(z_1-1) w_1^3\sum_{i=2}^5 f_{3,i}z^i\\&+w_1^4\sum_{i=2}^6 f_{4,i}z^i
+w_1^5\xi_1\sum_{i=3}^5 f_{5,i}z^i+w_1^6\xi_1^2\sum_{i=3}^5 f_{6,i}z^i+w_1^7\xi_1^3 f_{7,4}z^4+w_1^8\xi_1^4 f_{8,4}z^4
\end{split}
\ee
$g$ has similar structure, and we will not expand the details. Note that the divisor $z=0$ and the local patch $(z,w_1)$ still exist. In the patch $(z,w_1)$, we can choose $\xi_1=1$, so that $z=z_1$. Then it is easy to rewrite $f$ as a function only of $z$ and $w_1$.

Then in the patch $(z,w_1)$, we blow up a point $z=2$, $w_1=0$, which is a generic point on divisor $w_1=0$. After the blow up, we assign new coordinates $z_2,w_2$ and $\xi_2$:
\be
z-2=(z_2-2)\xi_2\ ,\ w=w_2\xi_2\label{blp2}
\ee

\begin{figure}
\centering
\includegraphics[height=5cm]{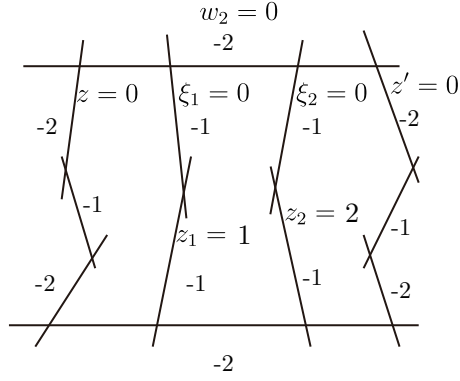}
\caption[E]{\footnotesize A semi-toric generalized $dP_7$, which is constructed by blowing up two points on the divisor $w=0$ on the toric base given in Figure~\ref{f:2Dtoric}.} \label{f:gdP7}
\end{figure}

We draw the geometry of the base after this blow up in Figure~\ref{f:gdP7}.

With a similar argument, after this blow up, $f$ now takes the form of 
\be
\bsp
f&=(z_1-1)^4(z_2-2)^4+f_1(z_1-1)^3 (z_2-2)^3 w_2 z+(z_1-1)^2 (z_2-2)^2 w_2^2 z\sum_{i=0}^2 f_{2,i}z^i
\\&+(z_1-1) (z_2-2) w_2^3 z^2\sum_{i=0}^2 f_{3,i}z^i+w_2^4 z^2\sum_{i=0}^4 f_{4,i}z^i+w_2^5\xi_1\xi_2 z^3\sum_{i=0}^2 f_{5,i}z^i
\\&+w_2^6\xi_1^2\xi_2^2 z^3\sum_{i=0}^2 f_{6,i}z^i+w_2^7\xi_1^3\xi_2^3 f_{7,4}z^4+w_2^8\xi_1^4\xi_2^4 f_{8,4}z^4\label{fgdP7}
\end{split}
\ee

$g$ has the similar structure. Note that the shape of the Newton polytopes for $f$ and $g$ has become a rhombus from a triangle.

\begin{figure}
\centering
\includegraphics[height=6cm]{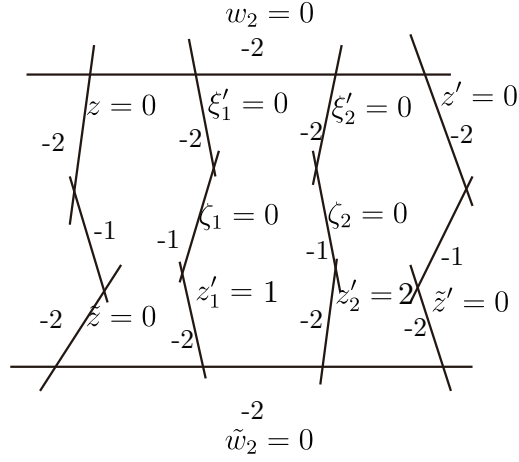}
\caption[E]{\footnotesize The local equations of divisors on $gdP_{9st}$.} \label{f:gdP9st}
\end{figure}

Finally, to get the $gdP_{9st}$, we need to blow up the points $z_1=1$, $\xi_1=0$ and $z_2=2$, $\xi_2=0$. After the blow-ups, we introduce new coordinates $z_1',\xi_1',\zeta_1$ and $z_2',\xi_2',\zeta_2$:
\be
z_1-1=(z_1'-1)\zeta_1\ ,\ \xi_1=\xi_1'\zeta_1\ ,\ (z_2-2)=(z_2'-2)\zeta_2\ ,\ \xi_2=\xi_2'\zeta_2
\ee
We draw the corresponding equations for the divisors on $gdP_{9st}$ in Figure~\ref{f:gdP9st}.

The requirement that $\zeta_1$ and $\zeta_2$ vanish to at least degree 4 in $f$ puts additional constraints on the coefficients in $f$. For example, the term $f_1(z_1-1)^3 (z_2-2)^3 w_2 z$ in (\ref{fgdP7}) has to vanish, since there is no way to have $\zeta_1^4\zeta_2^4$ in that term. Similar things happen for $g$. After this analysis, we divide $f$ by $\zeta_1^4\zeta_2^4$ and $g$ by $\zeta_1^6\zeta_2^6$. The final expression for $f$ and $g$ on $gdP_{9st}$ are:
\be
\bsp
f&=f_0(z_1'-1)^4 (z_2'-2)^4+f_1(z_1'-1)^3 (z_2'-2)^3 w_2^2 z\xi_1'\xi_2'+f_2(z_1'-1)^2 (z_2'-2)^2 w_2^4 z^2\xi_1'^2\xi_2'^2
\\&+f_3(z_1'-1) (z_2'-2) w_2^6 z^3\xi_1'^3\xi_2'^3+f_4 w_2^8 z^4\xi_1'^4\xi_2'^4
\end{split}
\ee
\be
\bsp
g&=g_0(z_1'-1)^6 (z_2'-2)^6+g_1(z_1'-1)^5 (z_2'-2)^5 w_2^2 z\xi_1'\xi_2'+g_2(z_1'-1)^4 (z_2'-2)^4 w_2^4 z^2\xi_1'^2\xi_2'^2
\\&+g_3(z_1'-1)^3 (z_2'-2)^3 w_2^6 z^3\xi_1'^3\xi_2'^3+g_4(z_1'-1)^2 (z_2'-2)^2 w_2^8 z^4\xi_1'^4\xi_2'^4\\
&+g_5(z_1'-1) (z_2'-2) w_2^{10} z^5\xi_1'^5\xi_2'^5+g_6 w_2^{12} z^6\xi_1'^6\xi_2'^6
\end{split}
\ee
We can restore the dependence on $z'$ by multiplying $z'^{8-m-n-p}$ factors to each term $z_1'^m z_2'^n z^p$ for $f$, and multiplying $z'^{12-m-n-p}$ factors to each term $z_1'^m z_2'^n z^p$ for $g$. The final expressions of $f$ and $g$ are:
\be
f=\sum_{i=0}^4 f_i\eta^i\tilde{\eta}^{4-i}\ ,\ g=\sum_{i=0}^6 g_i\eta^i\tilde{\eta}^{6-i}\ ,\ \eta\equiv w_2^2 z  z'\xi_1'\xi_2', \tilde{\eta}\equiv \tilde{w}_2^2 (z_1'-1) (z_2'-2)\tilde{z}\tilde{z}'.\label{fggdp}
\ee

Indeed, the monomials in $f$ and $g$ lie on a line. Moreover, the number of monomials in $f$ and $g$ are respectively 5 and 7, which exactly saturates the bound (\ref{FGbound}).

Similarly, we can explicitly compute the form of $f$ and $g$ for the other semi-toric bases in \cite{Martini-WT} with non-Higgsable U(1)s. Generally semi-toric bases are generated by blowing up Hirzebruch surfaces $\mathbb{F}_n$, in a way that the curves on it form chains between two specific curves $D_0$ and $D_\infty$, which correspond to the $(-n)$-curve $S$ and the $(+n)$-curve $\tilde{S}$ in the original $\mathbb{F}_n$. They are listed below, where $n_0$ and $n_\infty$ denotes the self-intersection number of $D_0$ and $D_\infty$. The chains are connected to $D_0$ at the front, and to $D_\infty$ at the end. Here $\eta$, $\xi$ and $\chi$ denote products of monomials, which are different from the notations in (\ref{fggdp}) and will not be specified.

We also list the rational sections in form of $(x,y,z)=(\lambda,\alpha,1)$.

Mordell-Weil rank $r=8$: $n_0=-2, n_\infty=-2$, $T=9$, $h^{1,1}=19$, $h^{2,1}=19$.

chain 1: $(-2,-1,-2)$

chain 2: $(-2,-1,-2)$

chain 3: $(-2,-1,-2)$

chain 4: $(-2,-1,-2)$
\bea
f&=&f_0\xi^4+f_1\xi^3\eta+f_2\xi^2\eta^2+f_3\xi\eta^3+f_4\eta^4\\
g&=&g_0\xi^6+g_1\xi^5\eta+g_2\xi^4\eta^2+g_3\xi^3\eta^3+g_4\xi^2\eta^4+g_5\xi\eta^5+g_6\eta^6
\eea
\bea
\lambda&=&\lambda_0\xi^2+\lambda_1\xi\eta+\lambda_2\eta^2\\
\alpha&=&\alpha_0\xi^3+\alpha_1\xi^2\eta+\alpha_2\xi\eta^2+\alpha_3\eta^3
\eea

Mordell-Weil rank $r=6$: $n_0=-2, n_\infty=-6$, $T=13$, $h^{1,1}=35$, $h^{2,1}=11$.

chain 1: $(-2,-1,-3,-1)$

chain 2: $(-2,-1,-3,-1)$

chain 3: $(-2,-1,-3,-1)$

chain 4: $(-2,-1,-3,-1)$
\bea
f&=&f_0 \xi^2\eta^2+f_1\xi\eta^3+f_2\eta^4\\
g&=&g_0\xi^4\eta^2+g_1\xi^3\eta^3+g_2\xi^2\eta^4+g_3\xi\eta^5+g_4\eta^6
\eea
\bea
\lambda&=&\lambda_0\xi\eta+\lambda_1\eta^2\\
\alpha&=&\alpha_0\xi^2\eta+\alpha_1\xi\eta^2+\alpha_2\eta^3
\eea

Mordell-Weil rank $r=6$: $n_0=-1, n_\infty=-2$, $T=10$, $h^{1,1}=24$, $h^{2,1}=12$.

chain 1: $(-3,-1,-2,-2)$

chain 2: $(-3,-1,-2,-2)$

chain 3: $(-3,-1,-2,-2)$
\bea
f&=&f_0 \xi^2\eta^2+f_1\xi\eta^5+f_2\eta^8\\
g&=&g_0\xi^4+g_1\xi^3\eta^3+g_2\xi^2\eta^6+g_3\xi^3\eta^9+g_4\eta^{12}
\eea
\bea
\lambda&=&\lambda_0\xi\eta+\lambda_1\eta^4\\
\alpha&=&\alpha_0\xi^2+\alpha_1\xi\eta^3+\alpha_2\eta^6
\eea

Mordell-Weil rank $r=5$: $n_0=-1, n_\infty=-8$, $T=16$, $h^{1,1}=51$, $h^{2,1}=3$.

chain 1: $(-3,-1,-2,-3,-2,-1)$

chain 2: $(-3,-1,-2,-3,-2,-1)$

chain 3: $(-3,-1,-2,-3,-2,-1)$
\bea
f&=&f_0\xi^2\eta^2\chi^2+f_1\xi^5\eta\chi^3\\
g&=&g_0\eta^4\chi^2+g_1\xi^3\eta^3\chi^3+g_2\xi^6\eta^2\chi^4
\eea
\bea
\lambda&=&\lambda_0\xi\eta\chi\\
\alpha&=&\alpha_0\eta^2\chi+\alpha_1\xi^3\eta\chi^2
\eea

Mordell-Weil rank $r=4$: $n_0=-2, n_\infty=-4$, $T=13$, $h^{1,1}=35$, $h^{2,1}=11$.

chain 1: $(-2,-2,-1,-4,-1)$

chain 2: $(-2,-2,-1,-4,-1)$

chain 3: $(-2,-2,-1,-4,-1)$
\bea
f&=&f_0\xi^4+f_1\xi^2\eta+f_2\eta^2\\
g&=&g_0\xi^6+g_1\xi^4\eta+g_2\xi^2\eta^2+g_3\eta^3
\eea
\bea
\lambda&=&\lambda_0\xi^2+\lambda_1\eta\\
\alpha&=&\alpha_0\xi^3+\alpha_1\xi\eta
\eea

Mordell-Weil rank $r=4$: $n_0=-6, n_\infty=-6$, $T=17$, $h^{1,1}=51$, $h^{2,1}=3$.

chain 1: $(-1,-3,-1,-3,-1)$

chain 2: $(-1,-3,-1,-3,-1)$

chain 3: $(-1,-3,-1,-3,-1)$

chain 4: $(-1,-3,-1,-3,-1)$
\bea
f&=&f_0\xi^2\eta^2\\
g&=&g_0\xi^4\eta^2+g_1\xi^3\eta^3+g_2\xi^2\eta^4
\eea
\bea
\lambda&=&\lambda_1\xi\eta\\
\alpha&=&\alpha_0\xi^2\eta+\alpha_1\xi\eta^2
\eea

Mordell-Weil rank $r=4$: $n_0=-1, n_\infty=-5$, $T=14$, $h^{1,1}=40$, $h^{2,1}=4$.

chain 1: $(-2,-1,-3,-1)$

chain 2: $(-4,-1,-2,-2,-3,-1)$

chain 3: $(-4,-1,-2,-2,-3,-1)$
\bea
f&=&f_0\xi^4+f_1\xi^2\eta^2\\
g&=&g_0\xi^6+g_1\xi^4\eta^2
\eea
\bea
\lambda&=&\lambda_0\xi^2\\
\alpha&=&\alpha_0\xi^3\textrm{ or }\xi^2\eta
\eea

Mordell-Weil rank $r=4$: $n_0=-1, n_\infty=-2$, $T=11$, $h^{1,1}=25$, $h^{2,1}=13$.

chain 1: $(-2,-1,-2)$

chain 2: $(-4,-1,-2,-2,-2)$

chain 3: $(-4,-1,-2,-2,-2)$
\bea
f&=&f_0\xi^4+f_1\xi^2\eta^2+f_2\eta^4\\
g&=&g_0\xi^6+g_1\xi^4\eta^2+g_2\xi^2\eta^4+g_3\eta^6
\eea
\bea
\lambda&=&\lambda_0\xi^2+\lambda_1\eta^2\\
\alpha&=&\alpha_0\xi^3+\alpha_1\xi\eta^2\textrm{ or }\alpha_0\xi^2\eta+\alpha_1\eta^3
\eea

Mordell-Weil rank $r=3$: $n_0=-4, n_\infty=-8$, $T=19$, $h^{1,1}=62$, $h^{2,1}=2$.

chain 1: $(-1,-4,-1,-2,-3,-2,-1)$

chain 2: $(-1,-4,-1,-2,-3,-2,-1)$

chain 3: $(-1,-4,-1,-2,-3,-2,-1)$

\bea
f&=&f_0\xi^2+f_1\xi\eta^2\\
g&=&g_0\xi^3+g_1\xi^2\eta^2
\eea
\bea
\lambda&=&\lambda_0\xi\\
\alpha&=&\alpha_0\xi\eta
\eea

Mordell-Weil rank $r=2$: $n_0=-2, n_\infty=-6$, $T=18$, $h^{1,1}=46$, $h^{2,1}=10$.

chain 1: $(-2,-1,-3,-1)$

chain 2: $(-2,-2,-2,-1,-6,-1,-3,-1)$

chain 3: $(-2,-2,-2,-1,-6,-1,-3,-1)$
\bea
f&=&f_0\xi^4+f_1\xi^2\eta\\
g&=&g_0\xi^6+g_1\xi^4\eta+g_2\xi^2\eta^2
\eea
\bea
\lambda&=&\lambda_0\xi^2\\
\alpha&=&\alpha_0\xi^3+\alpha_1\xi\eta
\eea

Mordell-Weil rank $r=2$: $n_0=-5, n_\infty=-6$, $T=21$, $h^{1,1}=61$, $h^{2,1}=1$.

chain 1: $(-1,-3,-1,-3,-1)$

chain 2: $(-1,-3,-2,-2,-1,-6,-1,-3,-1)$

chain 3: $(-1,-3,-2,-2,-1,-6,-1,-3,-1)$

\bea
f&=&f_0\xi^2\eta^2\\
g&=&g_0\xi^2+g_1\xi^4\eta^6
\eea
\bea
\lambda&=&\lambda_0\eta\xi\\
\alpha&=&\alpha_0\xi+\alpha_1\eta^2\xi^3
\eea

Mordell-Weil rank $r=2$: $n_0=-1, n_\infty=-2$, $T=12$, $h^{1,1}=24$, $h^{2,1}=12$.

chain 1: $(-2,-1,-2)$

chain 2: $(-3,-1,-2,-2)$

chain 3: $(-6,-1,-2,-2,-2,-2,-2)$

\bea
f&=&f_0\xi\eta+f_1\eta^4\\
g&=&g_0\xi^2+g_1\xi\eta^3+g_2\eta^6
\eea
\bea
\lambda&=&\lambda_0\eta^2\\
\alpha&=&\alpha_0\xi+\alpha_1\eta^3
\eea

Mordell-Weil rank $r=1$: $n_0=-2, n_\infty=-3$, $T=15$, $h^{1,1}=34$, $h^{2,1}=10$.

chain 1: $(-2,-1,-2)$

chain 2: $(-2,-2,-1,-4,-1)$

chain 3: $(-2,-2,-2,-2,-2,-1,-8,-1,-2)$

\bea
f&=&f_0\xi\eta+f_1\eta^4\\
g&=&g_0\xi\eta^3+g_1\eta^6
\eea
\bea
\lambda&=&\lambda_0\eta^2\\
\alpha&=&\alpha_0\eta^3
\eea

It would be interesting to construct the generators of the Mordell-Weil group for these models explicitly.

\section{A 3D base with non-Higgsable U(1)s}\label{s:3D}

Here we present a 3D base $B_3$ with non-Higgsable U(1)s, which is a higher dimensional analogy of $gdP_{9st}$ and saturates the bound (\ref{FGbound}) as well.

\begin{figure}
\centering
\includegraphics[height=6cm]{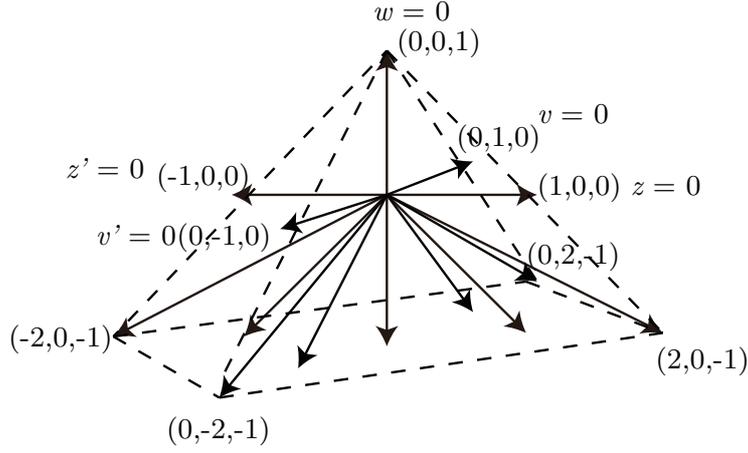}
\caption[E]{\footnotesize The toric rays of 3D toric variety $B_{3t}$ with toric rays $\{(0,0,1),(0,1,0),(1,0,0),(0,-1,0)$, $(-1,0,0)$, $(2,0,-1),(0,2,-1),(0,-2,-1),(-2,0,-1),(1,0,-1),(0,1,-1),(0,-1,-1),(-1,0,-1),(0,0,-1)\}$} \label{f:3Dtoric}
\end{figure}

\begin{figure}
\centering
\includegraphics[height=5cm]{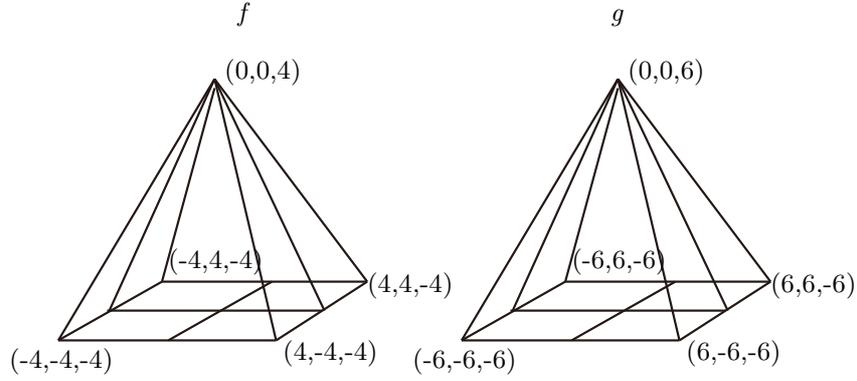}
\caption[E]{\footnotesize The set of monomials in $f$ and $g$, for the 3D toric variety $B_{3t}$ described in Figure~\ref{f:3Dtoric}} \label{f:3Dtoricfg}
\end{figure}

We start from a toric base $B_{3t}$ with toric rays shown in Figure~\ref{f:3Dtoric}. The sets of monomials in $f$ and $g$ are shown in Figure~\ref{f:3Dtoricfg}. Note that the assignment of 3D cones in the lower half space $z<0$ is not fully specified, hence there is some arbitrariness in defining this base geometry. The only invariant is the convex hull of 3D toric fan in Figure~\ref{f:3Dtoric} and the Newton polytopes for $f$ and $g$.  

In the local patches $(w,v,z)$, $(w,v,z')$, $(w,v',z)$ and $(w,v,z')$, $f$ and $g$ can be generally written as
\be
\bsp
f&=\sum_{i=0}^8\sum_{j=0}^8 f_{0,i,j}z^i z'^{8-i}v^j v'^{8-j}+w\sum_{i=1}^7\sum_{j=1}^7 f_{1,i,j}z^i z'^{8-i}v^j v'^{8-j}+w^2\sum_{i=1}^7\sum_{i=1}^7 f_{2,i,j}z^i z'^{8-i}v^j v'^{8-j}
\\&+w^3\sum_{i=2}^6\sum_{j=2}^6 f_{3,i,j}z^i z'^{8-i}+w^4\sum_{i=2}^6\sum_{j=2}^6 f_{4,i,j}z^i z'^{8-i}v^j v'^{8-j}
+w^5\sum_{i=3}^5\sum_{j=3}^5 f_{5,i,j}z^i z'^{8-i}v^j v'^{8-j}
\\&+w^6\sum_{i=3}^5\sum_{j=3}^5 f_{6,i,j}z^i z'^{8-i}v^j v'^{8-j}+w^7 f_{7,4,4}z^4 z'^4 v^4 v'^4+w^8 f_{8,4}z^4 z'^4 v^4 v'^4.
\end{split}
\ee

\be
\bsp
&g=\sum_{i=0}^{12}\sum_{j=0}^{12} g_{0,i,j}z^i z'^{12-i}v^j v'^{12-j}+w\sum_{i=1}^{11}\sum_{j=1}^{11} g_{1,i,j}z^i z'^{12-i}v^j v'^{12-j}+w^2\sum_{i=1}^{11}\sum_{j=1}^{11} g_{2,i,j}z^i z'^{12-i}v^j v'^{12-j}
\\&+w^3\sum_{i=2}^{10}\sum_{j=2}^{10} g_{3,i,j}z^i z'^{12-i}v^j v'^{12-j}+w^4\sum_{i=2}^{10}\sum_{j=2}^{10} g_{4,i,j}z^i z'^{12-i}v^j v'^{12-j}+w^5\sum_{i=3}^9\sum_{j=3}^{9} g_{5,i}z^i z'^{12-i,j}v^j v'^{12-j}
\\&+w^6\sum_{i=3}^9\sum_{j=3}^{9} g_{6,i,j}z^i z'^{12-i}v^j v'^{12-j}+w^7\sum_{i=4}^8\sum_{j=4}^{8} g_{7,i,j}z^i z'^{12-i}v^j v'^{12-j}+w^8\sum_{i=4}^8\sum_{j=4}^{8} g_{8,i,j}z^i z'^{12-i}v^j v'^{12-j}
\\&+w^9\sum_{i=5}^7\sum_{j=5}^{7} g_{9,i,j}z^i z'^{12-i}v^j v'^{12-j}+w^{10}\sum_{i=5}^7\sum_{j=5}^{7} g_{10,i,j}z^i z'^{12-i}v^j v'^{12-j}
\\&+w^{11} g_{11,6,6} z^6 z'^6 v^6 v'^6+w^{12}g_{12,6,6} z^6 z'^6 v^6 v'^6
\end{split}
\ee

\begin{figure}
\centering
\includegraphics[height=4cm]{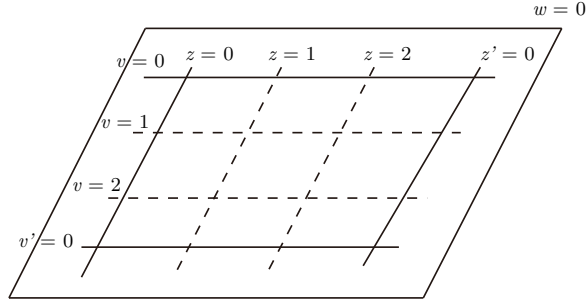}
\caption[E]{\footnotesize The series of blow-up on the divisor $w=0$ in the 3D toric variety $B_{3t}$. The dashed lines denote the curves to be blown up.} \label{f:3Dblp}
\end{figure}

Similar to the 2D case, we set $v'=1$, $z'=1$ for simplicity and we can restore them at last by multiplying the correct factors. Then we do the following 4 blow-ups (see Figure~\ref{f:3Dblp}):

(1) blow up the curve $z-1=w=0$, $z-1\rightarrow (z_1-1)\xi_1$, $w\rightarrow w_1\xi_1$.

(2) blow up the curve $z-2=w_1=0$, $z-2\rightarrow (z_2-2)\xi_2$, $w_1\rightarrow w_2\xi_2$.

(3) blow up the curve $v-1=w_2=0$, $v-1\rightarrow (v_1-1)\xi_3$, $w_2\rightarrow w_3\xi_3$.

(4) blow up the curve $v-2=w_3=0$, $v-2\rightarrow (v_2-2)\xi_4$, $w_3\rightarrow w_4\xi_4$.

After the process, $f$ becomes
\be
\bsp
f&=f_0(z_1-1)^4(z_2-2)^4(v_1-1)^4(v_2-2)^4+f_1 w_4 vz (z_1-1)^3(z_2-2)^3(v_1-1)^3(v_2-2)^3\\&+ w_4^2 vz (z_1-1)^2(z_2-2)^2(v_1-1)^2(v_2-2)^2 \sum_{i=0}^2\sum_{j=0}^2 f_{2,i,j}z^i v^j\\
&+w_4^3 v^2 z^2 (z_1-1)(z_2-2)(v_1-1)(v_2-2) \sum_{i=0}^2\sum_{j=0}^2 f_{3,i,j}z^i v^j+
w_4^4 v^2 z^2 \sum_{i=0}^4\sum_{j=0}^4 f_{4,i,j}z^i v^j\\
&+w_4^5 v^3 z^3\xi_1\xi_2\xi_3\xi_4 \sum_{i=0}^2\sum_{j=0}^2 f_{5,i,j}z^i v^j+w_4^6 v^3 z^3\xi_1^2\xi_2^2\xi_3^2\xi_4^2 \sum_{i=0}^2\sum_{j=0}^2 f_{6,i,j}z^i v^j\\
&+f_7 w_4^7 v^4 z^4\xi_1^3\xi_2^3\xi_3^3\xi_4^3+f_8 w_4^8 v^4 z^4\xi_1^4\xi_2^4\xi_3^4\xi_4^4
\end{split}
\ee
There is a similar expression for $g$.

Finally, we do the following 4 blow-ups.

(5) blow up the curve $z_1-1=\xi_1=0$, $z_1-1\rightarrow (z_1'-1)\zeta_1$, $\xi_1\rightarrow \xi_1'\zeta_1$.

(6) blow up the curve $z_2-2=\xi_2=0$, $z_2-2\rightarrow (z_2'-2)\zeta_2$, $\xi_2\rightarrow \xi_2'\zeta_2$.

(7) blow up the curve $v_1-1=\xi_3=0$, $v_1-1\rightarrow (v_1'-1)\zeta_3$, $\xi_3\rightarrow \xi_3'\zeta_3$.

(8) blow up the curve $v_2-2=\xi_4=0$, $v_2-2\rightarrow (v_2'-2)\zeta_4$, $\xi_4\rightarrow \xi_4'\zeta_4$.

After this process, we bring back $z'$ and $v'$ in a similar way as the 2D base. Finally, in the local patch near $w_4=0$, $f$ and $g$ can be written as
\be
f=\sum_{i=0}^4 f_i\eta^i\ ,\ g=\sum_{i=0}^6 g_i\eta^i\ ,\ \eta\equiv w_4^2 z z' v v'\xi_1'\xi_2'\xi_3'\xi_4'.
\ee
Similarly, there is another copy of this configuration on the opposite side, the local expression for $f$ and $g$ on that side can be written as

\be
f=\sum_{i=0}^4 f_{4-i}\tilde{\eta}^i\ ,\ g=\sum_{i=0}^6 g_{6-i}\tilde{\eta}^i\ ,\ \tilde{\eta}\equiv \tilde{w}_4^2 \tilde{z} \tilde{z}' \tilde{v} \tilde{v}' (z_1'-1)(z_2'-2)(v_1'-1)(v_2'-2).
\ee

The complete expressions of $f$ and $g$ are
\be
f=\sum_{i=0}^4 f_i\eta^i\tilde{\eta}^{4-i}\ ,\ g=\sum_{i=0}^6 g_i\eta^i\tilde{\eta}^{6-i}.
\ee

Because the form of $f$ and $g$ are the same as $gdP_{9st}$, we expect the Mordell-Weil rank to be 8.

This geometry resembles a K3 fibration over $\mathbb{P}^1$, with degenerate fibers $\eta=0$ and $\tilde{\eta}=0$. But they are not semi-stable degenerations \cite{Kulikov}, since the multiplicities of divisor $w_4$ and $\tilde{w}_4$ are 2, not 1. So we need to include more general degenerations that are not in Kulikov's list \cite{Kulikov}.

\begin{figure}
\centering
\includegraphics[height=5cm]{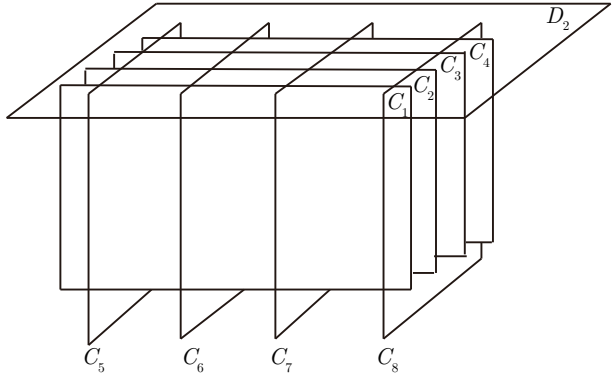}
\caption[E]{\footnotesize The irreducible components of divisor $D$ on the $B_3$, which is a degenerate K3 with vanishing canonical class.} \label{f:degK3}
\end{figure}

Here we explicitly check that the divisor $D$ corresponding to $\tilde{\eta}$ has vanishing canonical class: $K_D=0$. The structure of the components in $D$ is shown in Figure~\ref{f:degK3}:
\be
D=2D_2+C_1+C_2+C_3+C_4+C_5+C_6+C_7+C_8.
\ee

We have the adjunction formula 
\be
K_D=(K_X+D)|_D,
\ee
so we need to prove
\be
(K_X+D)\cdot D\sim 0\textrm{ on the divisor $D$.}
\ee

The local geometry of the divisor $D_2$ is the same as the toric divisor corresponding to the ray $(0,0,-1)$ on the toric threefold $B_{3t}$, see Figure~\ref{f:3Dtoric}. Also, the local geometry of divisors $C_i(1\leq i\leq 8)$ is the same as the toric divisor $(0,\pm 1,-1)$ or $(\pm 1,0,-1)$ on $B_{3t}$. Hence, the surfaces $D_2$ is the Hirzebruch surface $\mathbb{F}_0$ with $N=K_{\mathbb{F}_0}$, and $C_i(1\leq i\leq 8)$ are all Hirzebruch surfaces $\mathbb{F}_2$ with normal bundle $N=K_{\mathbb{F}_2}$. We then conclude that
\be
K_X|_{D_2}=0\ ,\ K_X|_{C_i}=0 (1\leq i\leq 8).
\ee

Hence we only need to prove
\be
D\cdot D\sim 0.
\ee

\begin{figure}
\centering
\includegraphics[height=6cm]{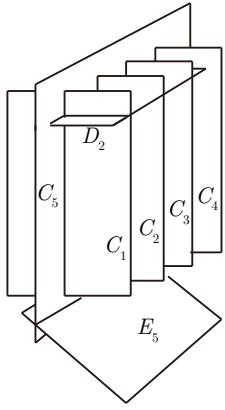}
\caption[E]{\footnotesize The local geometry of divisor $C_5$ in Figure~\ref{f:degK3}. Apart from $D_2$, $C_1$, $C_2$, $C_3$ and $C_4$, there is another divisor $E_5$ intersecting $C_5$.} \label{f:C5}
\end{figure}

First, notice that on the $\mathbb{F}_0$ surface $D_2$, the curves $D_2\cdot C_i (1\leq i\leq 4)$ all corresponds to same curve class. Similarly, the curves $D_2\cdot C_i (5\leq i\leq 8)$ all corresponds to same curve class. For $C_i$, we draw out its local geometry explicitly in Figure~\ref{f:C5}. On a divisor $C_i$, for any other $C_j$ and $C_k$ that intersects $C_i$, we have $C_i\cdot C_j\sim C_i\cdot C_k$. There is another divisor $E_5$ intersecting $C_5$, if we take $C_5$ to be the toric divisor $(1,0,-1)$ on $B_{3t}$, then this $E_5$ corresponds to the toric divisor $(2,0,-1)$. Similarly, because $C_5$ is a Hirzebruch surface $\mathbb{F}_2$, we have linear relations
\be
\bsp
C_5\cdot D_2&=C_5\cdot E_5-C_5\cdot C_1-C_5\cdot C_2\\
&=C_5\cdot E_5-2C_5\cdot C_1
\end{split}
\ee

Now we only need to compute $D_2^2$ and $C_i^2$. They can be computed by inspecting the toric divisors on $B_{3t}$. We take $(0,0,-1)$, $(1,0,-1)$, $(-1,0,-1)$, $(0,1,-1)$, $(0,-1,-1)$, $(2,0,-1)$ to be $D_2$, $C_5$, $C_6$, $C_1$, $C_2$ and $E_5$ respectively. Note that there is a toric linear relation
\be
D_2+C_5+C_6+C_1+C_2+X\sim 0,
\ee
where $X$ are divisors that do not intersect $D_2$. Hence we can compute
\be
D_2^2=D_2\cdot (-C_5-C_6-C_1-C_2)=-2D_2\cdot C_5-2D_2\cdot C_1.\label{D22}
\ee

Similarly near the divisor $C_5$, there is a linear relation
\be
C_5+E_5+D_2+C_1+C_2+Y\sim 0,
\ee
where $Y$ are divisors that do not intersect $C_5$. Hence we can compute
\be
C_5^2=C_5\cdot (-C_1-C_2-E_5-D_2)=-2D_2\cdot C_5-4C_1\cdot C_5.\label{C52}
\ee
Similarly,
\be
C_1^2=-2D_2\cdot C_1-4C_1\cdot C_5.
\ee

Finally, we can evaluate
\be
\bsp
D^2&=(2D_2+C_1+C_2+C_3+C_4+C_5+C_6+C_7+C_8)^2\\
&=4D_2^2+\sum_{i=1}^8 4D_2\cdot C_i+\sum_{i=1}^8 C_i^2+2(C_1+C_2+C_3+C_4)\cdot(C_5+C_6+C_7+C_8)\\
&=4D_2^2+16D_2\cdot C_1+16D_2\cdot C_5+\sum_{i=1}^8 C_i^2+2(C_1+C_2+C_3+C_4)\cdot(C_5+C_6+C_7+C_8)\\
&=8D_2\cdot C_1+8D_2\cdot C_5+\sum_{i=1}^8 C_i^2+32 C_1\cdot C_5\\
&=0
\end{split}
\ee
where we have plugged in (\ref{D22}), (\ref{C52}) and the equivalence between curve classes.

So we have proved that the divisor $D$ is indeed a degenerate K3. The relation $D^2=0$ implies that $D$ is a fiber.

\section{Conclusion}\label{s:conclusion}

In this paper, we developed formula (\ref{dh21g}) to count the change in Weierstrass moduli when a U(1) is tuned from the non-Higgsable phase of F-theory on an arbitrary base, if the Weierstrass form can be written in the Morrison-Park form (\ref{mp}). We argued that this formula should be exact for base point free line bundles $L$ parameterizing the Morrison-Park form. We have checked that this formula can even correctly take account of extra U(1) or SU(2) gauge groups in some special cases. Moreover, the formula (\ref{dh21g}) can be easily applied and computed for 3D toric bases. This is currectly the only tool of counting the number of neutral hypermultiplets in 4D F-theory setups with a U(1) gauge group, since there is no anomaly cancellation formula available for 4D F-theory.

We proposed that the choice $L=0$ in the Morrison-Park form corresponds to the minimal tuning of U(1) on a given base. Even when this choice is forbidden by the appearance of bad singularities, such as the $\mathbb{F}_{12}$ case, the formula (\ref{dh21L0}) provides a lower bound on the decrease in Weierstrass moduli. We proved the statement for 2D generalized del Pezzo (almost Fano) bases. However, we have not managed to formulate a complete proof of Conjecture 1, even for the generic 2D bases \cite{non-toric}. So it is still an open question to prove Conjecture 1 or find a counter example. 

Assuming Conjecture 1 holds, we derived stringent constraints on bases with non-Higgsable U(1)s. Even if Conjecture 1 does not hold, the statements about bases with non-Higgsable U(1)s may still hold, but they are technically much harder to prove. With the results, we confirmed that there is no non-Higgsable U(1) on toric bases. This result has been proved case by case for all the 2D toric bases \cite{mt-toric}, and it is useful to apply on the zoo of 3D toric bases \cite{Halverson-WT,MC}. Furthermore, we conjectured that a general $d$-dimensional base with non-Higgsable U(1)s should take the form of a resolution of a Calabi-Yau $(d-1)$-fold fibration over $\mathbb{P}^1$. It would be very interesting to extend the generalized Schoen construction in  \cite{FiberProduct} to degenerated K3 fibrations over $\mathbb{P}^1$. In particular, we need to include degenerate K3 fibers not in Kulikov's list\cite{Kulikov}. Using similar geometric techniques in Section~\ref{s:3D}, it may be possible to construct a base with non-Higgsable SU(3)$\times$SU(2)$\times$U(1), which can be useful in non-Higgsable standard model constructions \cite{ghst}. We leave this to future research.

Moreover, it is interesting to apply the techniques of Weierstrass moduli counting to the more general form of Weierstrass models with a rational section that give rise to U(1) charged matter with charge $q\geq 3$ \cite{Klevers-WT}, or Weierstrass models with two or three additional rational sections \cite{Cvetic:2013nia,Cvetic:2013qsa}. A related question is to algebraically construct Mordell-Weil generators if there are multiple of them, such as the examples listed in Section~\ref{s:2D}. We leave these questions to further research, too.

{\bf Acknowledgements}:

The author thanks Daniel Park and Washington Taylor for useful discussions and comments on the manuscript. The author also thanks David Morrison, Tom Rudelius and Cumrun Vafa for useful discussions. This research was supported by the DOE under contract 
\#DE-SC00012567.

\end{document}